\documentclass[aps,  prb,twocolumn,
  superscriptaddress,  floatfix,  10pt         
]{revtex4-2}

\usepackage{float}
\usepackage{graphicx}
\usepackage{amsmath,amssymb}
\usepackage{tikz}
\usepackage{braket}
\usepackage{pgfplots}
  \pgfplotsset{compat=1.17}
\usepackage[version=4]{mhchem}
\usepackage{physics}
\usepackage{subfigure}
\usepackage{enumitem}
\usepackage{mathrsfs}
\usepackage{mathtools}
\usepackage{subcaption}

\DeclarePairedDelimiter\fl{\lfloor}{\rfloor}

\usetikzlibrary{
  shapes.geometric,
  shadows,
  patterns,
  perspective,
  decorations,
  decorations.text
  }

\usetikzlibrary{decorations.markings,calc,decorations.pathreplacing}

\tikzset{
  baseline/.style = {line width=0.8pt, draw=black!70, line cap=round},
  tick/.style     = {line width=0.9pt, draw=black, line cap=round},
  site/.style     = {fill=black, draw=none},
  Jbond/.style    = {line width=1.4pt, draw=green!70!black, dotted, line cap=round},
  gbond/.style    = {line width=1.4pt, draw=cyan!80!black, line cap=round},
  Jlab/.style     = {font=\bfseries\footnotesize, text=green!60!black},
  glab/.style     = {font=\bfseries\footnotesize, text=cyan!80!black},
  upA/.style      = {-stealth, line width=0.5mm, draw=red!80!black},
  dnA/.style      = {-stealth, line width=0.5mm, draw=blue!80!black}
}

\newcommand{\nicechain}[4]{%
  \begin{scope}[shift={#1}]
    \def\dx{1.20}\def\Larr{1.00}\def\yoff{0.50}\def\rsite{0.12}\def\ticklen{0.17}
    \newcount\Np \Np=0
    \foreach \s [count=\j] in {#2} {\global\Np=\j}
    \newcount\N \N=\numexpr\Np-1\relax
    \ifnum\Np>1
      \draw[Jbond] ({1*\dx},0) -- ({2*\dx},0);
      \path ($({1*\dx},0)!0.5!({2*\dx},0)$) ++(0,0.20) node[Jlab]{J};
    \fi
    \ifnum\N>2
      \foreach \j in {2,...,\numexpr\N-1\relax}{
        \pgfmathtruncatemacro{\jp}{\j+1}
        \draw[gbond] ({\j*\dx},0) -- ({\jp*\dx},0);
        \path ($({\j*\dx},0)!0.5!({\jp*\dx},0)$) ++(0,0.20) node[glab]{g};
      }
    \fi
    \ifnum\Np>2
      \draw[Jbond] ({\N*\dx},0) -- ({\Np*\dx},0);
      \path ($({\N*\dx},0)!0.5!({\Np*\dx},0)$) ++(0,0.20) node[Jlab]{J};
    \fi
    \foreach \s [count=\j] in {#2}{
      \draw[tick] ({\j*\dx},-\ticklen) -- ({\j*\dx},\ticklen);
      \fill[site] ({\j*\dx},0) circle (\rsite);
      \ifnum\s=1
        \draw[upA] ({\j*\dx},-\yoff) -- ++(0,\Larr);
      \else
        \draw[dnA] ({\j*\dx},\yoff) -- ++(0,-\Larr);
      \fi
    }
  \end{scope}
}
\usepackage{xcolor}

\definecolor{linkcolor}{RGB}{0,0,255}      
\definecolor{citecolor}{RGB}{0,128,0}     
\definecolor{urlcolor}{RGB}{255,0,0}      

\usepackage{hyperref}
\hypersetup{
  colorlinks   = true,
  linkcolor    = linkcolor,
  citecolor    = citecolor,
  urlcolor     = urlcolor,
  linktoc      = all,
  pdfborder    = {0 0 0},
}

\definecolor{dy}{rgb}{0.9,0.9,0.4}
\definecolor{dr}{rgb}{0.95,0.65,0.55}
\definecolor{db}{rgb}{0.5,0.8,0.9}
\definecolor{dg}{rgb}{0.2,0.9,0.6}
\definecolor{BrickRed}{rgb}{0.8,0.3,0.3}
\definecolor{Navy}{rgb}{0.2,0.2,0.6}
\definecolor{DarkGreen}{rgb}{0.1,0.4,0.1}
\definecolor{phaseK}{HTML}{CC79A7}
\definecolor{phaseY}{HTML}{56B4E9}
\definecolor{phaseU}{HTML}{E69F00}
\definecolor{phaseF}{HTML}{23EB91}

\begin{document}

\title{Thermodynamics in a split Hilbert space: Quantum impurity at the edge of the Heisenberg chain}
\author{Abay Zhakenov}
\email{abay.zhakenov@rutgers.edu}
\author{Pradip Kattel}
\author{Natan Andrei}
\affiliation{Department of Physics and Astronomy, Center for Materials Theory, Rutgers University, Piscataway, New Jersey 08854, USA
}

\begin{abstract}
We study the isotropic spin-$\frac{1}{2}$ Heisenberg chain with a single edge-coupled impurity of arbitrary exchange strength $J$. The model exhibits four impurity phases. For antiferromagnetic couplings ($J>0$): a \textit{Kondo phase} at weak $J$, where the impurity is screened by many-body excitations and the impurity entropy decreases monotonically from $\ln 2$ at $T \to \infty$ to $0$ at $T \to 0$; and an \textit{antiferromagnetic bound-mode (ABM) phase} at strong $J$, where the impurity screened by an exponentially localized bound mode drives $S_{\mathrm{imp}}(T)$ nonmonotonically, with undershoots below zero at intermediate temperatures, while tending to $\ln 2$ as $T \to \infty$ and to $0$ as $T \to 0$. For ferromagnetic couplings ($J<0$): a \textit{local-moment (LM) phase} at weak $|J|$, where the impurity remains unscreened with $S_{\mathrm{imp}}\to \ln 2$ as $T \to 0$ but exhibits shallow undershoots at intermediate scales; and a \textit{ferromagnetic bound-mode (FBM) phase} at strong $|J|$, where $S_{\mathrm{imp}}=\ln 2$ in both UV and IR limits, yet develops an intermediate-temperature undershoot. We provide an analytic understanding of this behavior, showing that the undershoots originate from the fractionalization of the Hilbert space into several towers of states: for antiferromagnetic couplings this occurs only at strong $J$, driven by boundary-localized bound modes, while for ferromagnetic couplings undershoots occur for all $J<0$, becoming deeper with increasing $|J|$ and vanishing as $J \to 0^{-}$. These bound modes screen the impurity and reorganize the state structure, requiring a careful classification of states to be summed over to obtain a  consistent Thermodynamic Bethe Ansatz.  Incorporating the bound modes and edge states provides a complete analytic understanding of this phenomenon and yields closed expressions for the impurity contribution to free energy and entropy that are valid across all phases. These are checked and found to be in excellent agreement with tensor network and exact diagonalization results.
\end{abstract}

\maketitle

Quantum spin chains are paradigmatic models of strongly correlated matter, extensively studied both theoretically~\cite{Bethe1931,Hulthen1938,Orbach1958,Lieb1961,
YangYang1966I,YangYang1966II,YangYang1966III,Baxter1972,Takahashi1971,Takahashi1999,
LutherPeschel1975,Haldane1981,Haldane1983a,Haldane1983b,Affleck1989,Giamarchi2003,gaines2025spin}
 and experimentally \cite{Hutchings1969_KCuF3_AF_structure,Satija1980_KCuF3_spinwaves,Nagler1991_KCuF3_spin_dynamics,
Tennant1993_unbound_spinons_KCuF3_PRL,Tennant1995_KCuF3_continuum_PRB,
Buyers1986_Haldane_gap_CsNiCl3,Ma1992_NENP_long_lived,Regnault1994_NENP_PRB50,Dender1997_CuBenzoate_incomm,Hase1993_CuGeO3_spinPeierls,Regnault1996_CuGeO3_INS,Lake2005_QC_scaling_KCuF3_NatMat}
. Experimental realizations include a wide range of quasi-one-dimensional quantum magnets~\cite{scheie2022quantum,wang2018experimental,steiner1976theoretical,nagler1991spin} and are now routinely realized in quantum simulators such as ultracold atoms and Rydberg arrays, where couplings are tunable and defects can be engineered in a controlled way.~\cite{Salathe2015_PRX_DigitalQS,Monroe2021_RMP_TrappedIonSpinSims,Zhang2017_Nature_53QubitDPT,Zhang2009_PRA_NMR_IsingCriticality}. In these realizations, the chains are typically open and often at the edges the couplings can differ from those in the bulk due to the geometry or boundary conditions~\cite{chakhalian2004impurity,Aczel_Impurity-induced}.   The boundary sites act then as \emph{quantum impurities} and provide a basic example of a quantum spin impurity coupled to an interacting system.
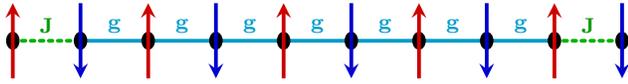
\begin{figure}[H]
    \centering
    \begin{tikzpicture}
      \begin{scope}[xscale=0.75]
        \nicechain{(0,0)}{1,-1,1,-1,1,-1,1,-1,1,-1}{}{ }
      \end{scope}
    \end{tikzpicture}
    \caption{Spin chain with $J$ (green, dotted) and $g$ (cyan, solid) impurity and bulk coupling, respectively.}
    \label{fig:spinchain}
\end{figure}

 \begin{figure}
 \includegraphics[width=\linewidth]{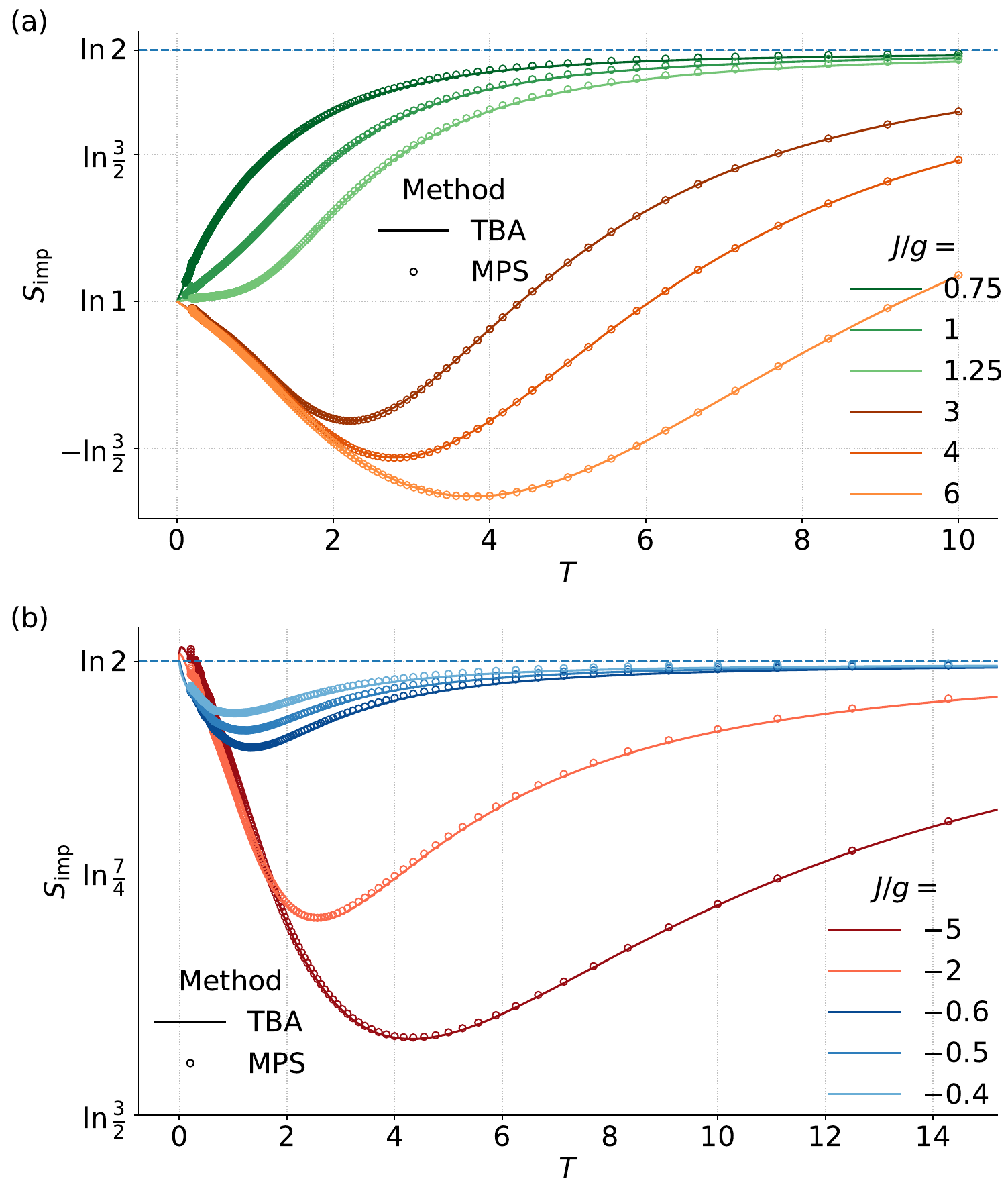}
    \caption{Impurity entropy $S_{\mathrm{imp}}(T)$ is shown in panels (a,b), computed via TBA (lines) and finite-$T$ tensor networks (dots). 
(a)\textbf{ Antiferromagnetic couplings}: the impurity goes from complete screening $S_{\mathrm{imp}}(T=0)=0$ to asymptotic freedom $S_{\mathrm{imp}}(T\to \infty)=\ln 2$. The Kondo phase (green) shows monotonic growth (consistent with conventional Kondo model), while the ABM phase (orange) is non-monotonic, with $S_{\mathrm{imp}}$ dipping below zero as the impurity binds into a localized singlet with the bound mode in $\mathcal{T}_{\rm BS}$, thereby removing that degree of freedom from the continuum. 
(b) \textbf{Ferromagnetic couplings}: the impurity remains unscreened with $S_{\mathrm{imp}}(0)=S_{\mathrm{imp}}(\infty)=\ln 2$ with dip in intermediate temperatures.
}
    \label{fig:AJimp}
\end{figure}

 Some of these systems are exactly solvable, allowing nonperturbative insight into many-body physics \cite{Bethe1931,franchini2017introduction,wang1997exact,frahm1997open,hu1998two,Kattel2024,kattel-edge2025}. A simple solvable system is the isotropic antiferromagnetic Heisenberg chain with a boundary impurity~\cite{wang1997exact,frahm1997open,laflorencie2008kondo,Kattel2024},
\begin{equation}\label{H}
  H= g\sum_{j=1}^{N-1} \vec{\sigma}_j \cdot \vec{\sigma}_{j+1}
   + J  \vec{\sigma}_1 \cdot \vec{\sigma}_0 ,
\end{equation}
where $g$ is the bulk exchange and $J$ the edge coupling (we will set $g=1$ for convenience of notation, restoring it when necessary for clarity. We consider the impurity only at one edge~\footnote{ We consider the effects of only one edge spin, assuming the chain is long enough so that the effects of the two edges are additive.}). The model is integrable for any value of the boundary coupling, allowing for a complete treatment of its properties.

 It was shown in Refs.~\cite{wang1997exact,Kattel2024} that the model exhibits several impurity phases: Kondo phase for weak antiferromagnetic coupling ($J\in(0,4/3)$), an antiferromagnetic bound mode (ABM) phase for strong antiferromagnetic coupling ($J\in(4/3,\infty)$), and two distinct ferromagnetic phases for $J<0$: a ferromagnetic bound mode (FBM) phase for $|J|>4/5$ and a local moment (LM) phase for $|J|<4/5$. In the Kondo phase, the impurity is screened by a many-body cloud of gapless bulk excitations, closely resembling the conventional Kondo effect, with the impurity critical exponents matching those of the usual single-channel Kondo problem~\cite{andrei1983solution,tsvelick1983exact,affleck1995conformal}. In the ABM phase, the impurity is effectively quenched by a single-particle bound mode that is exponentially localized near the impurity site. In the FBM phase, the impurity remains unscreened in the ground state but becomes screened at higher energies through the formation of a single-particle bound mode. Finally, in the LM phase, the impurity remains effectively unscreened across all energy scales, behaving as a free local spin.

Here, we proceed to study the finite-temperature thermodynamics of the model in all four boundary phases. In the Kondo phase, the impurity entropy behaves as expected: it approaches the asymptotically free value $S_{\mathrm{imp}}=\ln 2$ at high temperatures and decreases monotonically to the fully screened value $S_{\mathrm{imp}}=0$ at $T=0$~\cite{wang1997exact}, consistent with the spirit of the $g$-theorem~\cite{PhysRevLett.67.161,friedan2004boundary} (see Fig.\ref{fig:AJimp}). By contrast, in the other phases, the entropy shows a more unexpected behavior. In particular, $S_{\mathrm{imp}}(T)$ exhibits a nonmonotonic dip with negative value (see Fig.~\ref{fig:AJimp}) in intermediate temperatures. Note that similar results with a nonmonotonic dip were observed by some of us in non-interacting models analytically~\cite{kattel2024kondo,tang-twochannel} and in interacting systems numerically~\cite{kattel-edge2025}. We provide, for the first time, a complete analytic understanding of this phenomenon in an interacting spin chain. We show that the origin of the dip is the fractionalization of the Hilbert space into several towers of states, induced by the appearance of localized \emph{bound modes}. The bound modes reorganize the Hilbert space structure in all phases, and in certain phases they also screen the impurity~\cite{Kattel2024}, necessitating a careful classification of states in the evaluation of the free energy. We derive closed-form TBA expressions for the impurity contribution to free energy 
$F_{\mathrm{imp}}(T)$ and entropy $S_{\mathrm{imp}}(T)$, which can be evaluated analytically in the limits $T \to 0$ and $T \to \infty$, 
yielding results consistent with the expected values. At finite temperatures, we solve the equations numerically and benchmark the 
results against tensor network and exact diagonalization calculations 
(see Fig.\ref{fig:AJimp} and \cite{Supplemental}). Our framework extends directly to other integrable boundaries with localized modes (e.g.~XXZ, higher-spin SU(2), SU($n$) spin chains) and can be applied to field-theory settings~\cite{GNTBA}.

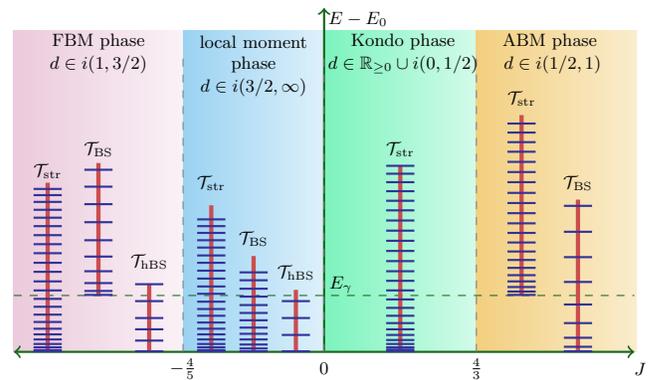
\begin{figure}
    \centering
    \begin{tikzpicture}[scale=0.75, every node/.style={scale=0.75}]

\shade[left color=phaseK!40, right color=phaseK!10] (-0.5,0) rectangle (2.5,5.7);
\shade[left color=phaseY!60, right color=phaseY!20] (2.5,0) rectangle (5,5.7);
\shade[left color=phaseF!60, right color=phaseF!20] (5,0) rectangle (7.7,5.7);
\shade[left color=phaseU!50, right color=phaseU!20] (7.7,0) rectangle (10.525,5.7);

\draw [thick,DarkGreen,<->] (-0.5,0)--(10.5,0);
\draw [thick,DarkGreen,<-] (5,6.1)--(5,0);
\node at (5.6,5.9) {\small $E - E_0$};
\node at (10.6,-0.3) {\small $J$};
\node at (2.5,-0.3) {\small $-\frac{4}{5}$};
\node at (5.0,-0.3) {\small $0$};
\node at (7.7,-0.3) {\small $\frac{4}{3}$};

\draw [dashed,DarkGreen] (-0.5,1.0)--(10.5,1.0);
\node at (5.3,1.2) {\small $E_\gamma$};

\foreach \x in {2.5,5.0,7.7}
  \draw[dashed,gray] (\x,0)--(\x,5.3);

\node at (1,5.5) {\small FBM phase};
\node at (1,5.1) {\small $d\in i(1,3/2)$};
\node at (3.75,5.5) {\small local moment};
\node at (3.75,5.1) {\small phase};
\node at (3.75,4.7) {\small $d \in i(3/2,\infty$)};
\node at (6.4,5.5) {\small Kondo phase};
\node at (6.4,5.1) {\small $d\in\mathbb{R}_{\geq 0} \cup i(0,1/2)$};
\node at (9,5.5)  {\small ABM phase};
\node at (9.05,5.1)  {\small $d\in i(1/2,1)$};

\node at (6.35,3.6) {$\mathcal{T}_\mathrm{str}$};
\node at (9.5,3) {$\mathcal{T}_\mathrm{BS}$};
\node at (0.1,3.2) {$\mathcal{T}_\mathrm{str}$};
\node at (3,3.0) {$\mathcal{T}_\mathrm{str}$};
\node at (1,3.6) {$\mathcal{T}_\mathrm{BS}$};
\node at (8.5,4.5) {$\mathcal{T}_\mathrm{str}$};
\node at (3.75,2) {$\mathcal{T}_\mathrm{BS}$};
\node at (1.9,1.6) {$\mathcal{T}_\mathrm{hBS}$};
\node at (4.5,1.4) {$\mathcal{T}_\mathrm{hBS}$};

\foreach \x/\ymin/\ymax in {0.1/0/3.0, 1/1/3.35, 1.9/0/1.2,
3/0/2.6, 3.75/0/1.7, 4.5/0/1.1,
    6.35/0/3.3,
    9.5/0/2.7, 8.5/1/4.2}
  \draw[ultra thick,BrickRed] (\x,\ymin)--(\x,\ymax);

\newcommand{\plotlevels}[2]{
  \foreach \e in {#2} {
    \draw[thick,Navy] (#1-0.25,\e)--(#1+0.25,\e);
  }
}

\plotlevels{6.35}{0.01,0.04,0.08,0.14,0.21,0.30,0.40,0.51,0.64,0.78,0.93,1.09,1.26,
1.43,1.61,1.79,1.97,2.15,2.33,2.51,2.68,2.85,3.01,3.16,3.30}

\plotlevels{9.5}{0.01,0.09,0.25,0.51,0.84,1.24,1.68,2.13,2.59}
\foreach \e in {0.01,0.07,0.15,0.25,0.36,0.49,0.63,0.78,0.94,1.11,
1.28,1.46,1.64,1.82,2.01,2.19,2.37,2.55,2.72,2.89,3.05}
  \draw[thick,Navy] (8.5-0.25,\e+1.0)--(8.5+0.25,\e+1.0);

\plotlevels{0.1}{0.01,0.03,0.08,0.14,0.22,0.31,0.41,0.53,0.66,0.80,0.94,1.09,
1.25,1.42,1.58,1.75,1.91,2.07,2.23,2.38,2.52,2.66,2.78,2.89}
\foreach \e in {0.01, 0.08, 0.22, 0.43, 0.69, 0.98, 1.30, 1.62, 1.93, 2.23}
  \draw[thick,Navy] (1-0.25,\e+1.0)--(1+0.25,\e+1.0);
\plotlevels{1.9}{0.01,0.2, 0.4,0.60, 0.9, 1.2}

\plotlevels{3}{0.01,0.03,0.08,0.14,0.22,0.31,0.41,0.52,0.64,0.76,0.89,
1.03,1.16,1.30,1.44,1.58,1.71,1.85,1.98,2.11,2.23,2.35}
\plotlevels{3.75}{0.01,0.06,0.13,0.22,0.32,0.43,0.56,0.69,0.83,0.98,1.12,1.27,1.41}
\plotlevels{4.5}{0.01,0.3,0.60, 0.9}

\end{tikzpicture}

    \caption{Phase diagram for Hamiltonian~\eqref{H} with an odd number of bulk sites. 
The spectrum evolves with the relative boundary coupling $J$. 
In the Kondo phase, the impurity is screened via the Kondo effect, and only the string tower is present. 
In the antiferromagnetic bound-mode (ABM) phase, a boundary-string (BS) tower emerges, describing localized screening by the bound mode, while the string tower is gapped by energy $E_\gamma$ [Eq.~\eqref{Egamma}]. 
In the ferromagnetic regimes, the spectrum comprises three towers: bulk strings, BS, and higher-order boundary strings (hBS). 
In the FBM, the BS tower is initially gapped, but as ferromagnetic coupling weakens, the gap closes, yielding a fourfold degenerate ground state.
}
    \label{fig:PD1}
\end{figure}

To obtain a full classification of the states, we analyze the solutions of the BA equations of this model~\cite{Supplemental}: 
\begin{equation}
\begin{aligned}\label{BAE}
    \left(\frac{\mu_j-i/2}{\mu_j+i/2}\right)^{2N} \frac{\mu_j-d-i/2}{\mu_j-d+i/2}\frac{\mu_j+d-i/2}{\mu_j+d+i/2}\\
    =\prod_{l\neq j}^M \frac{(\mu_j-\mu_l-i)(\mu_j+\mu_l-i)}{(\mu_j-\mu_l+i)(\mu_j+\mu_l+i)} 
    \end{aligned}
\end{equation}
with $M$ corresponding to the magnetization of the solution to be $S_z = \frac{N+1}{2}-M$, with the impurity parameter $d= d(J) =\sqrt{1/J-1} \in \mathbb{R}_{\geq 0} \cup i \mathbb{R}_{> 0}$.

In the \textit{ Kondo phase},  $J\in(0,4/3)$, the impurity parameter takes real values as well as imaginary values $d=i\gamma$ with $\gamma \in (0,1/2)$.  Within this regime, the Bethe ansatz equations  admit solutions of the form of strings given explicitly by:
\begin{equation}\label{strings}
\left\{\mu_{\alpha,j}^{(n)}\right\}_{j=1}^n
=
\left\{\Lambda_\alpha + \frac{i}{2}(n+1 - 2j)
 \middle|  j = 1,\dots,n\right\},
\end{equation}
where $\Lambda_\alpha \in \mathbb{R}$.
These string solutions span the entire Hilbert space, with each set of roots $\{\left\{\mu_{\alpha,j}\right\}_{j=1}^n\}$ corresponding to an eigenstate of the Hamiltonian. The ground state (GS) in particular is characterized by a configuration consisting of $M = (N+1)/2$ distinct 1-strings (real roots), denoted as $\mu_\alpha^{(1)}$ for $\alpha = 1, \dots, M$. Higher-energy states arise from different string configurations corresponding to collective spin-$1/2$ excitations known as spinons.

All these states collectively form a string \textit{tower of states} $\mathcal{T}_\mathrm{str}$ (see Fig.~\ref{fig:PD1}). Here we define a \textit{tower} as the set of states interconnected by spinon excitations, equivalently described as transitions between different string configurations. The lowest state of a tower we call its \textit{base state}. Each excited state within the tower thus represents a base state modified by the presence of spinon excitations.

This definition becomes particularly useful when examining the other three phases. As the impurity parameter becomes purely imaginary, $d = i\gamma$ with $\gamma > 1/2$, new classes of purely imaginary solutions of the Bethe Ansatz equations appear that are not of the form of the conventional string structure given by Eq.~\eqref{strings}. These include a boundary root (fundamental boundary string) $\mu_\gamma = i(\gamma - 1/2)$ and associated (higher-order) boundary $m$-strings defined by $\mu_{\gamma,l}^{(m)} = \mu_\gamma - i l$, where $l = 0, \dots, m$ and $m = \fl{\gamma + 1/2}$  (where $\lfloor\cdot\rfloor$ is the floor function)\cite{Wang2015}. These roots, being purely imaginary, correspond to exponentially localized bound modes.

In particular, in the \textit{ABM phase}, the impurity coupling becomes dominant $J>4/3$ and the impurity is screened by the boundary localized bound mode. This edge localization of the screening effect is the result of the ground state being described by the purely imaginary boundary root $\mu_\gamma$ in addition to a sea of 1-strings $\mu_\alpha^{(1)}$. The spinon excitations of the GS form a tower of states $\mathcal{T}_\mathrm{BS}$. These states, unlike those of $\mathcal{T}_\mathrm{str}$ contain the boundary root $\mu_\gamma$. The removal of this root causes the backflow of the 1-strings and unscreens the impurity, costing the energy increase of
\begin{equation}\label{Egamma}
    E_\gamma = \frac{2\pi g}{|\sin(\pi \gamma)|}
\end{equation}
Adding spinons on top of this base state forms the tower of states $\mathcal{T}_\mathrm{str}$ consisting only of various string and hole configurations and lacking the boundary root $\mu_\gamma$. This naturally splits the space of eigenstates of the Hamiltonian \eqref{H} for $J>4/3$ into two towers: $\mathcal{T}_\mathrm{BS}$ --- containing the boundary root $\mu_\gamma$, and $\mathcal{T}_\mathrm{str}$ --- not containing it with the latter being gapped by $E_\gamma$. In ~\cite{Supplemental}, we verify the existence of the two towers numerically via the exact diagonalization method. 

As the impurity parameter $\gamma$ is further increased  ($\gamma>1$), one explores the other two phases: \textit{the Ferromagnetic Bound Mode (FBM)} and \textit{the local moment} \textit{phases}. Both the string tower $\mathcal{T}_\mathrm{str}$ and the boundary root tower $\mathcal{T}_\mathrm{BS}$ are present there. However, unlike the Kondo and ABM phases, there are additional allowed solutions with the boundary string. The states containing the boundary string form the third tower - $\mathcal{T}_\mathrm{hBS}$. The base state of this tower corresponds to a singlet configuration of an unscreened impurity and a bulk spinon. Together with a complementary triplet base state of the tower $\mathcal{T}_\mathrm{str}$, they form a 4-degenerate ground state in the FBM phase corresponding to 4 possible states formed by the unscreened impurity and a bulk spinon. The boundary root tower $\mathcal{T}_\mathrm{BM}$ starts higher in spectrum with a singlet ($M_z=0$) base state gapped by $E_\gamma$ \eqref{Egamma} and corresponds to the screened impurity. In the local moment phase, all three towers start at the same energy level. So, in these two phases, the Hilbert space consists of three towers with the boundary root tower being gapped in the FBM phase by $E_\gamma$.

\paragraph*{The Thermodynamics:}
Having determined the Hilbert space structure, we proceed to  the study of impurity thermodynamics. 
For a chain with $N$ sites and open boundary conditions, we write
$
Z_0 = e^{-\beta F_{\rm bulk}(T)} \mathfrak g(T),
$
where $F_{\rm bulk}(T)$ is the extensive bulk free energy $(F_{\rm bulk}\propto N)$ and $\mathfrak g(T)$ is the universal $O(1)$ boundary factor \cite{pozsgay2010mathcal,he2024exact}. Introducing an impurity modifies the partition function to $Z = Z_0 \mathfrak g_{\rm imp}(T)$, so the impurity free energy gets the form $
F_{\rm imp}(T) = -\frac{1}{\beta}\ln \left(\frac{Z}{Z_0}\right) = -\frac{1}{\beta}\ln \mathfrak g_{\rm imp}(T).
$ Because we focus on the impurity contribution, we do not evaluate the open-edge factor $\mathfrak g(T)$; it can be obtained from \cite{Supplemental,pozsgay2010mathcal,he2024exact}. The impurity entropy is then $
S_{\rm imp}(T) = -\partial_T F_{\rm imp}(T)
= \ln \mathfrak g_{\rm imp}(T) + T \partial_T \ln \mathfrak g_{\rm imp}(T).
$

In the Kondo phase, the impurity is screened in the GS and gets asymptotically unscreened as $T\to \infty$, impurity's residual entropy grows from $S_{\rm{imp}}(T=0)=0$ to $S_{\rm{imp}}(T=\infty)=\ln{2}$. As the Hilbert space consists of only a single tower, there are no boundary excitations, and no other scales in the system. This suggests a monotonous growth of the residual entropy $S_{\rm{imp}}(T)$ with the temperature $T$. In this case, the conventional Thermodynamic Bethe Ansatz treatment allows us to express the impurity part of the free energy up to a GS energy shift as 
\begin{equation}
    F_{\rm{imp}} = -\frac{T}{4} \sum_{\upsilon = \pm} \int \frac{\ln\left[1 + \eta_1(\mu)\right]}{\cosh(\pi(\mu + \upsilon d))}  \dd\mu
\end{equation}
where $\eta_n$ is the hole/particle density ratio for $n$-strings given by Eq.~\eqref{strings}. These satisfy the coupled integral Thermodynamic Bethe Ansatz equations (TBA) (for details see \cite{Supplemental}) 
\begin{equation}
    \ln \eta_n(\lambda) = -\frac{2\pi}{T} \frac{\delta_{n,1}}{\cosh(\pi\lambda)}+\sum_{\upsilon=\pm 1}G\ln\left[ 1+\eta_{n+\upsilon} \right].\nonumber
\end{equation}
where the functional $G$ is defined as $Gf(\lambda)=\int \rm{d}\nu \frac{1}{2\cosh\pi(\lambda-\nu)} f(\nu)$ and the boundary conditions for the TBA recurssion relations are $\eta_0(\lambda)=0$ and
	\begin{equation}
		\lim_{n\to \infty}\left\{[n+1]\ln(1+\eta_n)-[n]\ln(1+\eta_{n+1})\right\}=-\frac{\mu h}{T}.\nonumber
	\end{equation}
    where $h$ is the applied global magnetic field and $\mu$ is the magnetic moment and the operator $[n]$ is defined as $[n]f(\nu)=\int \rm{d}\lambda \frac{1}{\pi} \frac{n/2}{\left(n/2\right)^2 + (\nu-\lambda)^2}f(\lambda)$. Analytic solutions for the TBA equations governing $\eta_n(T)$ are not available at general temperatures, but can be obtained in the asymptotic limits $T \to 0$ and $T \to \infty$. Therefore, to capture the full crossover behavior, we solve the TBA equations numerically \cite{Supplemental}. The resulting impurity entropy, shown in Fig.~\ref{fig:AJimp} (shades of green lines), confirms the expected asymptotic limits and yields a controlled description of the complete crossover from ultraviolet to infrared behavior across all temperatures.

However, the traditional TBA approach fails in the other phases, and a more careful analysis is needed. In the other three phases, the Hilbert space is split into several towers of excitations. As the free energy comes from these independent towers, the partition function shall be written as a sum of their contributions: $Z=e^{-\beta F} = \sum_{k\in\{\mathrm{towers}\}} Z_k=\sum_{k\in\{\mathrm{towers}\}} e^{-\beta F_k}$. The free energy expressions of the towers differ only in the impurity parts (see \cite{Supplemental} for details). This allows us to reorganize the partition function expression as 
\begin{equation}
    Z=e^{-\beta F_0}\sum_{k\in\{\mathrm{towers}\}} e^{-\beta F_k^{\mathrm{imp}}}= e^{-\beta F_0} e^{-\beta F^{\mathrm{imp}}} ,
\end{equation}
where the impurity part of the free energy becomes simply a weighted sum of the free energies of each of the towers \begin{equation}\label{Fimp}
    e^{-\beta F^{\mathrm{imp}}}\equiv  \sum_{k\in\{\mathrm{towers}\}} e^{-\beta F_k^{\mathrm{imp}}} .\end{equation}
This is the key idea that allows us to get correct thermodynamic results.

In the ABM phase, with a strong impurity coupling $J>4/3$, a new scale appears -- $E_\gamma$, the energy of an exponentially localized singlet screening the impurity. In the regime $g \ll T \ll E_\gamma$, the impurity forms a singlet with the boundary localized bound mode, thereby freezing out its local degrees of freedom. This results in $0\geq S_\mathrm{imp}(g\ll T\ll E_\gamma)\geq -\ln 2$. This behavior is most transparent in the limiting case $J\to \infty$, where the first site of the system locks into a singlet with the impurity and the system becomes effectively of $N-1$ sites. This analysis shows that in this case the impurity entropy should be a non-monotonous curve starting at zero, with a dip at $T\propto E_\gamma$ and then rising to $\ln 2$ as $T \to \infty$. Taking into account the
contributions of the towers $\mathcal{T}_\mathrm{str}$ and $\mathcal{T}_\mathrm{BS}$ (see \cite{Supplemental} for derivation)
\begin{equation*}
\begin{aligned}
F^{\rm{imp}}_{\mathrm{str}} &= E_\gamma-\frac{T}{4} \sum_{\upsilon = \pm} \int \frac{\ln\left[1 + \eta_2(\mu)\right]}{\cosh(\pi(\mu + \upsilon  i(\gamma-\frac{1}{2})) )}  \dd\mu\\
     &+\frac{T}{4} \sum_{\upsilon = \pm} \int \frac{\ln\left[1 + \eta_1(\mu)\right]}{ \cosh(\pi(\mu + \upsilon i(1-\gamma)))}  \dd\mu\\
F^{\rm{imp}}_{\mathrm{BS}} &=   \frac{T}{4} \sum_{\upsilon = \pm} \int  \frac{\ln\left[1 + \eta_1(\mu)\right]}{\cosh(\pi(\mu + \upsilon i(1-\gamma)))}  \dd\mu
\end{aligned}
\end{equation*}
and then solving the TBA equations numerically, one gets exactly this behavior (see various shades of orange curves in fig.~\ref{fig:AJimp}).
The string tower $\mathcal{T}_\mathrm{str}$, having a gap, starts to influence the impurity free energy at temperatures comparable to the gap energy $T\sim E_\gamma$. This causes the system first to explore the low-lying bound mode states from the boundary root tower $\mathcal{T}_\mathrm{BS}$, which brings negative entropy contribution as these states correspond to the bulk excitations of the locally screened impurity.
This non-monotonous behavior of the impurity coincides with an independent result obtained from a finite temperature tensor network calculation performed in the ITensor library~\cite{verstraete2004matrix,zwolak2004mixed,ITensor1,Itensor2} and ED results \cite{Supplemental}. At infinite temperature, the entropy contribution of each tower is proportional to its dimensionality as $S_k^{\mathrm{str}}(T\to \infty)= \ln ( \frac{Z_{k}}{Z_0})$, where $k$ represents a tower: either the string $\mathcal{T}_\mathrm{str}$ or the boundary root $\mathcal{T}_\mathrm{BS}$ in this case. From the free energy expressions we obtain the infinite temperature entropies \cite{Supplemental}   $S^\mathrm{imp}_{\mathrm{str}}(T\to \infty)=\ln 3/2$ and $S^\mathrm{imp}_{\mathrm{BS}}(T\to \infty)=\ln 1/2$ yielding that the string and boundary root towers contain $\dim \mathcal{T}_\mathrm{str}=3/2 \cdot 2^N$ and $\dim \mathcal{T}_\mathrm{BS}=1/2 \cdot 2^N$ states correspondingly totaling the dimensionality of the full Hilbert space $\dim \mathcal{H} = 2^{N+1}$.

In the ferromagnetic phases, the impurity is unscreened in both the ground state and at $T \to \infty$, so $S_{\rm imp}(0)=S_{\rm imp}(\infty)=\ln 2$. The contributions of each of the towers $\mathcal{T}_\mathrm{str}$, $\mathcal{T}_\mathrm{BM}$, and $\mathcal{T}_\mathrm{BS}$ are
\begin{align}
F^{\rm{imp}}_{\mathrm{str}}&= - \frac{T}{4} \sum_{\upsilon = \pm} \int \mathrm{d}\lambda 
\left[
\frac{\ln \left(1 + \eta_{\fl{2\gamma}+1}(\lambda)\right)}
     {\cosh \left(\pi\left(\lambda + \frac{i\upsilon}{2}(2\gamma - \lfloor 2\gamma \rfloor)\right)\right)}
\right. \nonumber\\
&\hspace{4.5em}\left.
- \frac{\ln \left(1 + \eta_{\lfloor 2\gamma \rfloor}(\lambda)\right)}
     {\cosh \left(\pi\left(\lambda + \frac{i\upsilon}{2}(\fl{2\gamma} - 2\gamma+1)\right)\right)}
\right], \nonumber\\
F^{\rm{imp}}_{\mathrm{BS}} &=\frac{T}{4} \sum_{\upsilon = \pm} \int \mathrm{d}\lambda 
\left[
\frac{\ln \left(1 + \eta_{\fl{2\gamma} - 1}(\lambda)\right)}
     {\cosh \left(\pi\left(\lambda + \frac{i\upsilon}{2}(2\gamma - \lfloor 2\gamma \rfloor)\right)\right)}
\right. \nonumber\\
&\hspace{2.5em}\left.
- \frac{\ln \left(1 + \eta_{\lfloor 2\gamma \rfloor - 2}(\lambda)\right)}
     {\cosh \left(\pi\left(\lambda + \frac{i\upsilon}{2}(\fl{2\gamma} - 2\gamma+1)\right)\right)}
\right]+E_\gamma, \nonumber\\
F^{\rm{imp}}_{\mathrm{hBS}} &= \frac{T}{4} \sum_{\upsilon = \pm} \int \mathrm{d}\lambda 
\left[
\frac{\ln \left(1 + \eta_{\lfloor 2\gamma \rfloor}(\lambda)\right)}
     {\cosh \left(\pi\left(\lambda + \frac{i\upsilon}{2}(\fl{2\gamma} - 2\gamma+1)\right)\right)}
\right. \nonumber\\
&\hspace{4.5em}\left.
+ \frac{\ln \left(1 + \eta_{\fl{2\gamma} - 1}(\lambda)\right)}
     {\cosh \left(\pi\left(\lambda + \frac{i\upsilon}{2}(2\gamma - \lfloor 2\gamma \rfloor)\right)\right)}
\right]. \nonumber
\end{align}
with $\gamma\in(1,3/2)$ in the FBM phase and $\gamma>3/2$ in the local Moment phase. Solving the TBA equation numerically for $\eta_n$, one gets non-monotonous impurity behavior for both of these phases (see Fig.~\ref{fig:AJimp}).
The dimensionality of the towers can be obtained from their infinite temperature entropy contributions \cite{Supplemental}. In the FBM phase, the Hilbert space splits into towers with dimensions: $\dim \mathcal{T}_\mathrm{str}=4/3 \cdot 2^N$, $\dim \mathcal{T}_\mathrm{BS}=1/2 \cdot 2^N$, and $\dim \mathcal{T}_\mathrm{hBS}=1/6 \cdot 2^N$.  In the local moment phase on the other hand their dimensionalities depend on $\gamma$: $\dim \mathcal{T}_\mathrm{str}=\frac{\fl{2\gamma}+2}{\fl{2\gamma}+1} \cdot 2^N$, $\dim \mathcal{T}_\mathrm{BS}=\frac{\fl{2\gamma}-1}{\fl{2\gamma}} \cdot 2^N$, and $\dim \mathcal{T}_\mathrm{hBS}=\frac{1}{\fl{2\gamma}(\fl{2\gamma}+1)} \cdot 2^N$. Thus, in the vanishing ferromagnetic coupling limit $J\to 0^-$ ($\gamma\to \infty$) the string tower and the boundary-root tower each occupy half of the Hilbert space $\dim \mathcal{T}_\mathrm{str}\to 2^N$, $\dim \mathcal{T}_\mathrm{BS}\to 2^N$ while the higher-order boundary-string tower contributes only a thermodynamically negligible fraction of states, $\dim T_\mathrm{hBS} / \dim T_\mathrm{str, BS} \to 0$.

\textit{In sum:} We present an exact, thermodynamically consistent classification of a boundary impurity in the isotropic Heisenberg chain in which boundary roots reorganize the Hilbert space into excitation towers $\mathcal{T}_{\rm str}$, $\mathcal{T}_{\rm BS}$, and $\mathcal{T}_{\rm hBS}$. The activation scale $E_\gamma=\frac{2\pi g}{|\sin(\pi\gamma)|}$ controls inter-tower mixing. The zero- and high-temperature limits are:
\begin{enumerate}
    \item \textit{Kondo} ($0<J<4/3$): $S_{\rm imp}(0)=0$ and $S_{\rm imp}(\infty)=\ln 2$; $S_{\rm imp}(T)$ increases monotonically with $T$.
    \item \textit{ABM} ($J>4/3$): $S_{\rm imp}(0)=0$, with a non-monotone dip for $g\ll T\ll E_\gamma$ (localized screening removes bulk degrees of freedom, yielding $0\ge S_{\rm imp}\gtrsim-\ln 2$), and $S_{\rm imp}(\infty)=\ln 2$.
    \item \textit{Ferromagnetic (FBM and local Moment, $\gamma>1$)}: the impurity is unscreened at both ends, $S_{\rm imp}(0)=\ln 2$ and $S_{\rm imp}(\infty)=\ln 2$; intermediate-$T$ shows non-monotone dips (vanishing as $\gamma \to \infty$ in the local-moment regime).
\end{enumerate}

Our tower-summed TBA  yields closed forms for $F_{\rm imp}(T)$ and $S_{\rm imp}(T)$ with the correct $T \to 0,\infty$ limits, in quantitative agreement with tensor network/ED benchmarks. Beyond this model, the framework generalizes—via reflection algebra—to XXZ, higher-spin SU(2), and SU($n$) chains (and field-theory limits), providing a unified recipe for impurities/defects that generate boundary-bound modes and sharp predictions for thermodynamic signatures accessible in cold-atom and Rydberg-array platforms.

Our results raise an important question: does the Hilbert-space fragmentation that alters thermodynamics also manifest in quantum dynamics and non-equilibrium settings? While we leave a full study to future work, in (supplemental material)~\cite{Supplemental}, we show that the impurity spectral function differs quantitatively across the four phases, suggesting that dynamical and non-equilibrium phenomena will likewise reflect this effect.

{\it Acknowledgments:} We thank P.~Pasnoori, P.~Azaria, and J.~H.~Pixley  for collaboration on related work and useful discussions, and C.~Rylands and Y.~Tang for helpful discussions.

\bibliography{ref}

\widetext
	\newpage
	\begin{center}
		\textbf{\large Supplementary Materials for `Thermodynamics in a split Hilbert space: Quantum impurity at the edge of the Heisenberg chain'}\\
		\vspace{0.4cm}
		 Abay Zhakenov,$^1$  Pradip Kattel,$^1$ and Natan Andrei$^1$\\
		\vspace{0.2cm}
		\small{$^1$\textit{Department of Physics and Astronomy, Center for Materials Theory, Rutgers University, Piscataway, New Jersey 08854, USA}}
	\end{center}
	
	\renewcommand{\thesection}{S.\arabic{section}}

	\setcounter{equation}{0} 
	\setcounter{page}{1}
	
	\renewcommand{\theequation}{S.\arabic{equation}}
    
In this Supplementary Material, we provide additional details supporting the results presented in the main text. We first present the full derivation of the thermodynamic Bethe Ansatz (TBA) equations in each phase, together with the expressions for the free energy contribution due to impurity. Numerical methods for solving the truncated TBA equations are described, along with convergence checks. Finally, we present exact diagonalization results showing the structure of towers in the antiferromagnetic bound mode phase and also discuss the finite temperature tensor network methods used for the computation of impurity entropy.

\section*{Derivation of the TBA equations}
\subsection{String tower}
The BAE is
\begin{equation}
    \left(\frac{\mu_j-i/2}{\mu_j+i/2}\right)^{2N} \frac{\mu_j-i/2-i\gamma}{\mu_j+i/2-i\gamma}\frac{\mu_j-i/2+i\gamma}{\mu_j+i/2+i\gamma}=\prod_{l\neq j}^M \frac{(\mu_j-\mu_l-i)(\mu_j+\mu_l-i)}{(\mu_j-\mu_l+i)(\mu_j+\mu_l+i)}  
\end{equation}

Writing it for $n$-strings and taking log one gets
\begin{equation}
\begin{aligned}\label{log BAE}
    (2N+1) \Theta_{n}(\lambda_j^n)+\Theta_{n}(\lambda_j^n-i\gamma)+\Theta_{n}(\lambda_j^n+i\gamma) = 2\pi J_j^n +\sum_{m=1}^{\infty} \sum_{i=1}^{\zeta_m}\left(\Theta_{nm}(\lambda_j^n - \lambda_i^m)+\Theta_{nm}(\lambda_j^n + \lambda_i^m)\right)
\end{aligned}
\end{equation}
with $J_j^n$ being distinct integers fixing the root positions of the $n$-strings.

Let's introduce the notations used above. The basic element of all these functions is 
\begin{equation}
    \Theta_n (x) = 2\arctan(\frac{2x}{n}) .
\end{equation}
The function $\Theta_{nm}(x)$ is defined through $\Theta_n$ as
\begin{equation}
    \begin{aligned}
        \Theta_{nm}(x)&=
        \begin{cases}
            \Theta_{|n-m|}(x)+2\Theta_{|n-m|+2}(x)+\ldots + 2\Theta_{n+m-2}(x)+\Theta_{n+m}(x) , & n\neq m\\
             2\Theta_{2}(x)+2\Theta_{4}(x)+\ldots + 2\Theta_{2n-2}(x)+\Theta_{2n}(x) , & n=m
        \end{cases} .
    \end{aligned}
\end{equation}
Taking derivative of the eq.~\eqref{log BAE} with respect to the roots position and defining
\begin{equation}
    \begin{aligned}
        \dv{J_n}{\mu}=\sigma_n(\mu)+\sigma_n^h(\mu)
    \end{aligned}
\end{equation}
we rewrite the BAE as
\begin{equation}\label{sigma_n^h}
    \sigma_n^h(\mu) = f_n^{\mathrm{str}}(\mu) - \sum_{m=1}^{\infty}A_{nm} \sigma_m(\mu)
\end{equation}
The notations used above are
\begin{equation}
    \begin{aligned}
        f_n^{\mathrm{strings}}(\mu)&=(2N+1) K_n(\mu)+K_{n}(\mu+i\gamma)+K_{n}(\mu-i\gamma) ,\\
        A_{nm}&=\left[|n-m|\right]+2\left[|n-m|+2\right]+\ldots+2\left[n+m-2\right]+\left[n+m\right]
    \end{aligned}
\end{equation}
with $K_n(\mu)$ defined as
\begin{equation}
    \begin{aligned}
        K_n(\mu) &\equiv \frac{1}{2\pi} \dv{\Theta_n}{\mu} =\frac{1}{\pi}\frac{(n/2)}{(n/2)^2+\mu^2}
    \end{aligned}
\end{equation}
and functional $\left[ n\right]$ introduced as convolution with $K_n$:
\begin{equation}
    \left[n\right] g(\mu)\equiv K_n * g(\mu) = \int \dd \lambda  K_n(\mu-\lambda)g(\lambda)
\end{equation}

The $m$-string located at position $\lambda^m$ has the energy $E_m(\lambda^m) = -4\pi K_m(\lambda^m)$. So, the total energy can be written as
\begin{equation}
    E=-4\pi \sum_{n=1}^\infty \sum_{j=1}^{\zeta_n} K_n(\lambda^n_j) = -2\pi \sum_{n=1}^\infty \int \dd \lambda  K_n(\lambda) \sigma_n(\lambda)
\end{equation}
The magnetization of a state with $M$ roots consisting of $\zeta_n$ $n$-strings is $S^z=(N+1)/2-M=(N+1)/2-\sum_{n=1}^{\infty}n\zeta_n$ which in thermodynamic limit reads as
\begin{equation}
    S^z = \frac{N+1}{2}-\frac{1}{2}\sum_{n=1}^\infty n\int \dd \lambda  \sigma_n(\lambda)
\end{equation}
Note that in the last two equations, when going into the thermodynamic limit, the sum was swapped by an integral with an additional factor of $1/2$. This is to take into account that the roots $\lambda$ and $-\lambda$ correspond to the same state, so as integrals go from $-\infty$ to $\infty$ we count the same root twice.

The string part of the free energy can be written as
\begin{equation}
    F_{\mathrm{strings}}=E-S^z H - T\mathcal{S}
\end{equation}
where $\mathcal{S}$ is the Yang-Yang entropy after Stirling approximation:
\begin{equation}
    \mathcal{S}=\frac{1}{2}\sum_{n=1}^\infty \int \dd \lambda  \left[ (\sigma_n+\sigma_n^h)\ln(\sigma_n+\sigma_n^h)-\sigma_n\ln\sigma_n - \sigma_n^h\ln\sigma_n^h \right]
\end{equation}
Combining $E+MH$ as
\begin{equation}
    E+M H = \frac{1}{2}\sum_{n=1}^\infty \int \dd \lambda   g_n(\lambda) \sigma_n(\lambda)
\end{equation}
where we introduced $g_n(\lambda)=nH-4\pi K_n(\lambda)$, one can write the free energy as
\begin{equation}\label{F}
    F_{\mathrm{strings}}=\frac{1}{2}\sum_{n=1}^\infty \int \dd \lambda   \left[ g_n \sigma_n - T\sigma_n \ln\left[1+\frac{\sigma_n^h}{\sigma_n}\right] - T\sigma_n^h \ln\left[1+\frac{\sigma_n}{\sigma_n^h}\right] \right]-\frac{N+1}{2}H
\end{equation}
Varying the free energy and using $\delta \sigma_n^h = -\sum_{m=1}^\infty A_{nm}\delta \sigma_m$  from \eqref{sigma_n^h} we get
\begin{equation}
    g_n-T\ln\left[1+\frac{\sigma_n^h}{\sigma_n}\right]+T\sum_{m=1}^\infty A_{nm}\ln\left[1+\frac{\sigma_m}{\sigma_m^h}\right]=0
\end{equation}
or, introducing $\eta_n = \sigma_n^h/\sigma_n$, one can write the mimimum free energy condition as
\begin{equation}\label{pre TBA}
\ln\left[1+\eta_n(\lambda)\right]=\frac{g_n(\lambda)}{T}+\sum_{m=1}^\infty A_{nm} \ln\left[1+\eta^{-1}_m(\lambda)\right] .
\end{equation}
Defining the functional $G$ acting by convolution with $1/2\cosh(\pi \lambda)$
\begin{equation}
    Gf(\lambda) = \int \dd \mu   \frac{1}{2\cosh\pi (\lambda-\mu)}f(\mu)
\end{equation}
we can apply $\delta_{nm}-(\delta_{n+1,m}+\delta_{n-1,m})G$ onto \eqref{pre TBA} we get the TBA equation
\[
\ln\left[1+\eta_n(\lambda)\right]-G\ln\left[ 1+\eta_{n+1} \right]-G\ln\left[ 1+\eta_{n-1} \right] = -\frac{2\pi}{T} \frac{\delta_{n,1}}{\cosh(\pi\lambda)}+\ln\left[1+\eta^{-1}_n(\lambda)\right],
\]
or simpler
\begin{equation}\label{TBA}
    \ln \eta_n(\lambda) = -\frac{2\pi}{T} \frac{\delta_{n,1}}{\cosh(\pi\lambda)}+G\ln\left[ 1+\eta_{n+1} \right]+G\ln\left[ 1+\eta_{n-1} \right].
\end{equation}

Additionally one can notice that $[n]K_l(\mu) = K_n*K_l(\mu) = K_{n+l}(\mu)$ which gives in infinite $n$ limit $n\rightarrow\infty$
\begin{equation*}
    \lim_{n\rightarrow \infty} [n+1]\ln(1+\eta_n(\mu))=\frac{1}{T}\left( nH-4\pi K_{2n+1}(\mu) \right)+\sum A_{2n+1,m} \ln[1+\eta^{-1}_m(\mu)].
\end{equation*}
and
\begin{equation*}
    \lim_{n\rightarrow \infty} [n]\ln(1+\eta_{n+1}(\mu))=\frac{1}{T}\left( (n+1)H-4\pi K_{2n+1}(\mu) \right)+\sum A_{2n+1,m} \ln[1+\eta^{-1}_m(\mu)],
\end{equation*}
or 
\begin{equation}\label{TBA limit}
    \lim_{n\rightarrow \infty} \left\{[n+1]\ln(1+\eta_n(\mu))-[n]\ln(1+\eta_{n+1}(\mu))\right\}=-\frac{H}{T}.
\end{equation}
After performing the saddle point approximation \eqref{pre TBA} one gets 
\begin{equation}
    \begin{aligned}\label{F2}       F_{\mathrm{saddle}}&= -\frac{T}{2}\sum_n \int \dd \lambda f_n(\lambda) \ln \left[1+\eta^{-1}_n(\lambda)\right] - H\frac{N+1}{2}\\
        &=-\frac{T}{2}\sum_n \int \dd \lambda \left( (2N+1) K_n(\lambda) +K_{n}(\lambda+ i \gamma)+K_{n}(\lambda- i \gamma) \right) \ln \left[1+\eta^{-1}_n(\lambda)\right] - H\frac{N+1}{2}
    \end{aligned}
\end{equation}
Acting with the functional $G$ on the equation \eqref{pre TBA} we get
\begin{equation}\label{trick}
    G\left[ \ln(1+\eta_m(\lambda))-\frac{g_m(\lambda)}{T} \right]=\sum_n Y_{mn}*\ln(1+\eta^{-1}_n(\lambda))
\end{equation}
where we introduced a function $Y_{mn}(\mu)$
\begin{equation}
    Y_{m,n}(\mu)\equiv G*A_{mn}(\mu)=\sum_{l=1}^{\mathrm{min}(n,m)} K_{n+m+1-2l}(\mu)
\end{equation}
Using that $Y_{1n}(x)=K_n(x)$, the bulk term in \eqref{F2} can be rewritten as
\[
\sum_n \int \dd \lambda K_n (\lambda) \ln(1+\eta^{-1}_n(\lambda)) = \sum_n Y_{1,n} * \ln(1+\eta^{-1}_n (0)) =G\left[ \ln(1+\eta_{1}(\lambda))-\frac{g_{1}(\lambda)}{T} \right].
\]
Finally, the free energy can be written as \eqref{F2} as
\begin{equation}\label{F str}
    \begin{aligned}
        F_{\mathrm{saddle}}&=\frac{1}{2}\int \dd \lambda  \left\{ \frac{2N+1}{2\cosh \pi \lambda}\left[ g_{1}(\lambda)-T\ln(1+\eta_{1}(\lambda)) \right] \right\}-H\frac{N+1}{2}\\
        &-\frac{T}{2}\sum_n \int \dd \lambda \left\{K_{n}(\lambda+i\gamma)+K_{n}(\lambda-i\gamma)\right\} \ln \left[1+\eta^{-1}_n(\lambda)\right]\\
        &=F_{0}+F^{\mathrm{imp}}_{\mathrm{str}}
    \end{aligned}
\end{equation}
where we defined by $F_{0}$ the free energy part of the bulk, and the last term is the impurity part:
\begin{equation*}
    F^{\mathrm{imp}}_{\mathrm{str}}=-\frac{H}{2}-\frac{T}{2}\sum_n \int \dd \lambda \left\{K_{n}(\lambda+i\gamma)+K_{n}(\lambda-i\gamma)\right\} \ln \left[1+\eta^{-1}_n(\lambda)\right]
\end{equation*}

Performing the saddle point approximation we get only one term of the total partition function. An extra term comes due to the density of states Jacobian and quadratic fluctuations around the
saddle. The exact result for the partition function is
\begin{equation}
    Z= e^{-\beta F_{\text{saddle}}} 
    \frac{\det(\mathbf{1}-\widehat{\mathcal Q}^{-})}{\det(\mathbf{1}-\widehat{\mathcal Q}^{+})},
\end{equation}
so that the total spin free energy is
\begin{equation}\label{true F}
    F = F_{\text{saddle}}
    -T\left[\log\det(\mathbf{1}-\widehat{\mathcal Q}^{-})
    -\log\det(\mathbf{1}-\widehat{\mathcal Q}^{+})\right].
\end{equation}
The operators $\widehat{\mathcal Q}^{\pm}$ act on sequences of functions $f = \{f_m(\mu)\}_{m \ge 1}$ defined for positive rapidities $\mu > 0$ according to

$$
(\widehat{\mathcal Q}^{\pm} f)_n(\lambda) = \sum_{m=1}^\infty \int_0^\infty \frac{d\mu}{2\pi} \, \frac{f_m(\mu) \, \bigl[a_{nm}(\lambda-\mu) \pm a_{nm}(\lambda+\mu)\bigr]}{1+\eta_m(\mu)},
$$

with the kernel $a_{nm}(x)$ expressed as the derivative of the scattering phase:

$$
a_{nm}(x) = \frac{1}{2\pi} \frac{d}{dx} \Theta_{nm}(x).
$$

The quantity $F_{\mathrm{saddle}}$ represents the saddle-point free energy, incorporating both bulk and impurity contributions derived earlier. The accompanying ratio of determinants arises from the nontrivial normalization of the functional integral and yields an $O(1)$ contribution to the free energy, independent of the impurity. The factor $1/(1+\eta_m(\mu))$ reflects the equilibrium occupation ratios $\eta_m(\mu) = \sigma_m^h(\mu)/\sigma_m(\mu)$ as determined by the TBA equations \eqref{TBA}. By taking the ratio of partition functions with and without the impurity, one isolates the impurity-specific contribution without needing to evaluate the two Fredholm determinants explicitly.

In what follows, for simplicity, we will consider only the zero magnetic field case $H=0$ so that the impurity free energy of string solutions is
\begin{equation}\label{F imp str}
    F^{\mathrm{imp}}_{\mathrm{str}}=-\frac{T}{2}\sum_n \int \dd \lambda \left\{K_{n}(\lambda+i\gamma)+K_{n}(\lambda-i\gamma)\right\} \ln \left[1+\eta^{-1}_n(\lambda)\right]
\end{equation}

The identity \eqref{trick} can be analytically continued to the strip in complex line $\lambda\pm i \delta$ with $|\delta|<1/2$:
\begin{equation}\label{trick C}
    \sum_{\pm}G\left[ \ln(1+\eta_m(\lambda\pm i\delta))-\frac{g_m(\lambda\pm i\delta)}{T} \right]=\sum_{\pm}\sum_n Y_{mn}*\ln(1+\eta^{-1}_n(\lambda\pm i\delta))
\end{equation}

To use it for the \eqref{F imp str}, we can write $\gamma$ as the sum of the closest smaller (half-)integer and the remaining positive part: $\gamma=\frac{\fl{2\gamma}}{2}+\delta$ with $\delta\in\left[0,1/2\right)$. Then
\[
\begin{aligned}
    K_{n}(\lambda+i\gamma)+K_{n}(\lambda-i\gamma) = K_{n}\left(\lambda+i\frac{\fl{2\gamma}}{2}+i\delta\right)+K_{n}\left(\lambda-i\frac{\fl{2\gamma}}{2}-i\delta\right)\\
   =K_{n}\left(\lambda+i\frac{\fl{2\gamma}}{2}+i\delta\right)+K_{n}\left(\lambda+i\frac{\fl{2\gamma}}{2}-i\delta\right)-K_{n}\left(\lambda+i\frac{\fl{2\gamma}}{2}-i\delta\right)+\ldots\\
   -K_{n}\left(\lambda-i\frac{\fl{2\gamma}}{2}+i\delta\right)+K_{n}\left(\lambda-i\frac{\fl{2\gamma}}{2}+i\delta\right)+K_{n}\left(\lambda-i\frac{\fl{2\gamma}}{2}-i\delta\right)\\
   =K_{n+ \fl{2\gamma}}(\lambda\pm i \delta)+K_{n+ \fl{2\gamma}-2}(\lambda\pm i \delta)+\ldots + K_{n+2- \fl{2\gamma}}(\lambda\pm i \delta)+K_{n- \fl{2\gamma}}(\lambda\pm i \delta)\\
   -\left[K_{n+ \fl{2\gamma}-1}\left(\lambda\pm i \left(\tfrac{1}{2}-\delta\right)\right)+K_{n\pm \fl{2\gamma}-3}\left(\lambda\pm i\left(\tfrac{1}{2}-\delta\right)\right)+\ldots +K_{n+1- \fl{2\gamma}}\left(\lambda\pm i\left(\tfrac{1}{2}-\delta\right)\right)\right]\\
   =Y_{n,\fl{2\gamma}+1}(\lambda\pm i \delta)-Y_{n,\fl{2\gamma}}\left(\lambda\pm i\left(\tfrac{1}{2}-\delta\right)\right)
\end{aligned}
\]
So, \eqref{F imp str} becomes
\begin{equation}
    F^{\mathrm{imp}}_{\mathrm{str}}=-\frac{T}{2}\sum_n \int \dd \lambda \left\{Y_{n,\fl{2\gamma}+1}(\lambda\pm i \delta)-Y_{n,\fl{2\gamma}}(\lambda\pm i\left(\tfrac{1}{2}-\delta\right))\right\} \ln \left[1+\eta^{-1}_n(\lambda)\right]
\end{equation}
and using \eqref{trick C} we get
\begin{equation}
\begin{aligned}
    F^{\mathrm{imp}}_{\mathrm{str}}&=\frac{1}{2}G*\left[g_{\fl{2\gamma}+1}(\pm i \delta)-g_{\fl{2\gamma}}\left(\pm i \left(\tfrac{1}{2}-\delta\right)\right)\right]-\frac{T}{2}G*\left\{ \ln\left[1+\eta_{\fl{2\gamma}+1}(\pm i\delta)\right]-\ln\left[1+\eta_{\fl{2\gamma}}\left(\pm i \left(\tfrac{1}{2}-\delta\right)\right)\right] \right\}\\
    &=-2\pi G*\left(K_{2\gamma+1}(0)-K_{2\gamma-1}(0)\right) \\
    &-\frac{T}{2}\int \dd \lambda \sum_{v=\pm} G(\lambda+iv\delta) \ln\left[1+\eta_{\fl{2\gamma}+1}(\lambda)\right]+\frac{T}{2}\int \dd \lambda \sum_{v=\pm} G(\lambda+iv\left(\tfrac{1}{2}-\delta\right)) \ln\left[1+\eta_{\fl{2\gamma}}(\lambda)\right]
\end{aligned}
\end{equation}
The first term is the impurity part of the string tower base energy $E_{\mathrm{str}}$. For $\gamma<1/2$ it is
\begin{equation}
    E_{\mathrm{str}} = -2\pi G*\left(K_{1+2\gamma}(0)+K_{1-2\gamma}(0)\right) =- \int \dd \omega \frac{e^{-(1+\gamma)|\omega|}+e^{-(1-\gamma)|\omega|}}{1+e^{-|\omega|}}=\sum_{v=\pm}\left(\psi\left(\frac{1+v \gamma}{2}\right)-\psi\left(\frac{2+v\gamma}{2}\right)\right)
\end{equation}
where $\psi(x)$ is a digamma function. 
And for $\gamma>1/2$
\begin{equation}
    E_{\mathrm{str}} = -2\pi G*\left(K_{2\gamma+1}(0)-K_{2\gamma-1}(0)\right) = \int \dd \omega \frac{e^{-\gamma|\omega|}-e^{-(\gamma+1)|\omega|}}{1+e^{-|\omega|}}=2\psi\left(\frac{\gamma+1}{2}\right)-\psi\left(\frac{\gamma+2}{2}\right)-\psi\left(\frac{\gamma}{2}\right)
\end{equation}

At $T=0$ and $T=\infty$, the ratios $\eta_n(\lambda)$ become constant functions $\eta_n^{T=0,\infty}$ and the TBA equations \eqref{TBA} become algebraic:
\begin{equation}
\begin{aligned}
    \ln \eta_1^{T=0} &= -\infty\\
    \ln \eta_n^{T=0}  &=  \frac{1}{2}\ln[(1+\eta_{n-1}^{T=0} )(1+\eta_{n+1}^{T=0})]\\
         \lim_{n\rightarrow \infty} \frac{\ln \eta_n^{T=0} }{n} &= 0
\end{aligned}
\end{equation}
and
\begin{equation}
\begin{aligned}
    \ln \eta_1^{T=\infty} &= \frac{1}{2}\ln[(1+\eta_{2}^{T=\infty} )]\\
    \ln \eta_n^{T=\infty}  &=  \frac{1}{2}\ln[(1+\eta_{n-1}^{T=\infty} )(1+\eta_{n+1}^{T=\infty})]\\
         \lim_{n\rightarrow \infty} \frac{\ln \eta_n^{T=\infty} }{n} &= 0
\end{aligned}
\end{equation}
This fixes the ratios to be
\begin{equation}\label{eta const}
\begin{aligned}
    \eta_n^{T=0}&=n^2-1\\
    \eta_n^{T=\infty}&=(n+1)^2-1
\end{aligned}
\end{equation}
This results in the free energy at $T=0$ to be
\begin{equation}
    \begin{aligned}
        F^{\mathrm{imp}}_{\mathrm{str}}(T\to 0) &= E_\mathrm{str}-\frac{T}{2} \ln(\frac{1+\eta_{\fl{2\gamma}+1}^{T=0}}{1+\eta_{\fl{2\gamma}}^{T=0}})\\
        &=E_\mathrm{str}-T \mathcal{S}_\mathrm{str}^{\mathrm{imp}}(T=0)
    \end{aligned}
\end{equation}
with the string tower impurity entropy contribution  at $T=0$ being
\begin{equation}
    \mathcal{S}_\mathrm{str}^{\mathrm{imp}}(T=0)=\begin{cases}
        0 , & \gamma<1/2\\
        \ln(\frac{\fl{2\gamma}+1}{\fl{2\gamma}}) , & \gamma>1/2
    \end{cases}
\end{equation}
At $T=\infty$, \eqref{eta const} result into 
\begin{equation}
    \begin{aligned}
        \lim_{T\to \infty} \frac{F^{\mathrm{imp}}_{\mathrm{str}}(T) }{T} = -\frac{1}{2}\ln(\frac{1+\eta_{\fl{2\gamma}+1}^{T=\infty}}{1+\eta_{\fl{2\gamma}}^{T=\infty}})=-\mathcal{S}_\mathrm{str}^{\mathrm{imp}}(T=\infty)
    \end{aligned}
\end{equation}
with the string tower entropy at infinite temperature being 
\begin{equation}
    \mathcal{S}_\mathrm{str}^{\mathrm{imp}}(T=\infty)=\ln(\frac{\fl{2\gamma}+2}{\fl{2\gamma}+1})
\end{equation}

Additionally, one can calculate the bulk part of the saddle point entropy in \eqref{F str} at infinite temperature:
\begin{equation}
    \begin{aligned}
        \mathcal{S}_\mathrm{saddle}^{0}(T=\infty) = -\lim_{T\to \infty} \frac{F^{0}(T) }{T} = \frac{2N+1}{4}\ln(1+\eta_{1}^{T=\infty})=N \ln 2 + \frac{\ln 2}{2}
    \end{aligned}
\end{equation}
The actual bulk infinite temperature entropy of the bulk is the log of dimensionality of the Hilbert space of $N$ spin-1/2 sites: $\mathcal{S}^{0}(T=\infty) = N\log 2$. It shows that the saddle point value differs the correct one by $1/2 \ln 2$. Using \eqref{true F} one can see that this should be the value of the Fredholm determinants at infinite temperature.

\subsection{Boundary string}
Whenever $\gamma>1/2$ the BAE \eqref{BAE} allows for an additional solution $\mu=i(\gamma-1/2)$. Adding this root explicitly, the BAE \eqref{log BAE} for $n$-strings changes to
\begin{equation}
\begin{aligned}\label{log BAE gamma b 1}
    (2N+1) \Theta_n (\lambda_j^n) -\sum_{v=\pm}\Theta_{n}(\lambda_j^n+i v (\gamma-1))= 2\pi J_j^n +\sum_{m=1}^{\infty} \sum_{i=1}^{\zeta_m}\left(\Theta_{nm}(\lambda_j^n - \lambda_i^m)+\Theta_{nm}(\lambda_j^n + \lambda_i^m)\right)
\end{aligned}
\end{equation}
This modifies equation \eqref{sigma_n^h} relating the densities of holes and roots to:
\begin{equation}\label{sigma_n^h gamma}
    \sigma_n^h(\mu) = f_n^{BS}(\mu) - \sum_{m=1}^{\infty}A_{nm} \sigma_m(\mu)
\end{equation}
where the new function $f_n^{BS}(\mu)$ is defined as:
\begin{equation}
    \begin{aligned}
       f_n^{BS}(\mu)&=(2N+1)K_n(\mu)-\sum_{v=\pm}K_{n}(\mu+iv(\gamma-1))
    \end{aligned}
\end{equation}
and the free energy, similar to \eqref{F str}, is
\begin{equation}
    \begin{aligned}
        F_{\mathrm{BS}}&=F_{0}+4\pi K_1(i(\gamma-1/2))+\frac{T}{2}\sum_n \int \dd \lambda \sum_{v=\pm}K_{n}(\lambda+iv(\gamma-1)) \ln \left[1+\eta^{-1}_n(\lambda)\right]
    \end{aligned}
\end{equation}
here the second term is the energy of the added boundary root $\mu_\gamma=i(\gamma-1/2)$.
The boundary string tower has the same bulk part of the free energy $F_{0}$ and differs from the string tower only in the impurity contribution $F_{\mathrm{BS}}^\mathrm{imp}$. In this case it is
\begin{equation}\label{F BS imp}
    F_{\mathrm{BS}}^\mathrm{imp}=4\pi K_1(i(\gamma-1/2))+\frac{T}{2}\sum_n \int \dd \lambda \sum_{v=\pm}K_{n}(\lambda+iv(\gamma-1)) \ln \left[1+\eta^{-1}_n(\lambda)\right]
\end{equation}
Similarly to the string tower, for $\gamma>3/2$, it can be simplified to
\begin{equation}
\begin{aligned}
    F^{\mathrm{imp}}_{\mathrm{BS}}=&2\pi G*\left(K_{2\gamma-1}(0)-K_{2\gamma-3}(0)\right)-4\pi K_1(i(\gamma-1/2)) \\
    &+\frac{T}{2}\int \dd \lambda \sum_{v=\pm} G(\lambda+iv\delta) \ln\left[1+\eta_{\fl{2\gamma}-1}(\lambda)\right]-\frac{T}{2}\int \dd \lambda \sum_{v=\pm} G(\lambda+iv\left(\tfrac{1}{2}-\delta\right)) \ln\left[1+\eta_{\fl{2\gamma}-2}(\lambda)\right]
\end{aligned}
\end{equation}
with $\delta = \gamma - \frac{\fl{2\gamma}}{2}$ and the first term is the impurity part of the boundary string tower base energy $E_{\mathrm{BS}}$
\begin{equation}
    E_{\mathrm{BS}} = 2\pi G*\left(K_{2\gamma-1}(0)-K_{2\gamma-3}(0)\right)-4\pi K_1(i(\gamma-1/2)) = \int \dd \omega \frac{e^{-\gamma|\omega|}-e^{-(\gamma-1)|\omega|}}{1+e^{-|\omega|}}-4\pi K_1(i(\gamma-1/2))=E_{\mathrm{str}}
\end{equation}
which coincides with the string tower base energy $E_{\mathrm{str}}$ for the considered case of $\gamma>3/2$.
This changes when $\gamma \in (1/2,3/2)$. The boundary string impurity free energy in this case becomes 
\begin{equation}
\begin{aligned}
     F^{\mathrm{imp}}_{\mathrm{BS}}&=2\pi G*\left(K_{2\gamma-1}(0)+K_{3-2\gamma}(0)\right)+4\pi K_1(i(\gamma-1/2)) +\frac{T}{2}\int \dd \lambda \sum_{v=\pm} G(\lambda+iv(\gamma-1)) \ln\left[1+\eta_{1}(\lambda)\right]\\
    &=E_\mathrm{str}-\frac{2\pi}{\sin(\pi \gamma)}+\frac{T}{2}\int \dd \lambda \sum_{v=\pm} G(\lambda+iv(\gamma-1)) \ln\left[1+\eta_{1}(\lambda)\right]
\end{aligned}
\end{equation}
with the second term making the base energy of this tower lower than the string tower one for ABM phase $\gamma\in(1/2,1)$ and higher for the FBM phase $\gamma \in (1,3/2)$. 

Using \eqref{eta const}, the zero-temperature entropy contribution of this tower is
\begin{equation}
    \mathcal{S}_\mathrm{BS}^{\mathrm{imp}}(T=0)=\begin{cases}
        0 , & \gamma\in(1/2,3/2)\\
        \ln(\frac{\fl{2\gamma}-2}{\fl{2\gamma}-1}) , & \gamma>3/2
    \end{cases}
\end{equation}
Similarly, for $T\to \infty$
\begin{equation}
    \mathcal{S}_\mathrm{BS}^{\mathrm{imp}}(T\to \infty)=\begin{cases}
        \ln(1/2), & \gamma \in (1/2,3/2)\\
        \ln(\frac{\fl{2\gamma}-1}{\fl{2\gamma}}) , & \gamma>3/2
    \end{cases}
\end{equation}

\subsection{Higher-order boundary string}
Starting $\gamma>1$, the BAE \eqref{BAE} allows for the higher-order boundary strings: $\mu_{\gamma,l}^{(m)} = \mu_\gamma - i l$, where $l = 0, \dots, m$ and $m = \fl{\gamma + 1/2}$ \cite{Wang2015}. Adding them explicitly to the equation \eqref{BAE}, one gets a new relation for the density of holes: 
\begin{equation}\label{sigma_n^h gamma_a}
    \sigma_n^h(\mu) = f_n^{hBS}(\mu) - \sum_{m=1}^{\infty}A_{nm} \sigma_m(\mu)
\end{equation}
where the new function $f_n^{hBS}(\mu)$ is defined as:
\begin{equation}
    \begin{aligned}
       f_n^{hBS}(\mu)&=(2N+1)K_n(\mu)-\left[2\sum_{k=1}^{m}\sum_{v=\pm}K_{n}(\mu+iv(\gamma-k))+K_{n}(\mu+iv(\gamma-m-1))\right]
    \end{aligned}
\end{equation}
and the impurity  free energy ends up to be
\begin{equation}
    \begin{aligned}\label{F hBS imp}
    F_{\mathrm{hBS}}^\mathrm{imp}&=-4\pi \sum_{k=0}^{m} K_1(i(\gamma-1/2-k))\\
        &+\frac{T}{2}\sum_n \int \dd \lambda \sum_{v=\pm}\left[2\sum_{k=1}^{m}\sum_{v=\pm}K_{n}(\lambda+iv(\gamma-k))+K_{n}(\lambda+iv(\gamma-m-1))\right] \ln \left[1+\eta^{-1}_n(\lambda)\right]
    \end{aligned}
\end{equation}
here the first term is the energy of the added higher-order boundary string roots.
As in the previous two cases, after applying Takahashi identity \eqref{trick C} it simplifies to
\begin{equation}
F_{\mathrm{hBS}}^{\mathrm{imp}}= E_{\mathrm{str}}+ \frac{T}{4} \sum_{\upsilon = \pm} \int \mathrm{d}\lambda 
\left[
\frac{\ln \left(1 + \eta_{\lfloor 2\gamma \rfloor}(\lambda)\right)}
     {\cosh \left(\pi\left(\lambda + \frac{i\upsilon}{2}(\fl{2\gamma} - 2\gamma+1)\right)\right)}
+ \frac{\ln \left(1 + \eta_{\fl{2\gamma} - 1}(\lambda)\right)}
     {\cosh \left(\pi\left(\lambda + \frac{i\upsilon}{2}(2\gamma - \lfloor 2\gamma \rfloor)\right)\right)}
\right]. 
\end{equation}

Applying \eqref{eta const}, the zero-temperature entropy contribution of this tower is
\begin{equation}
    \mathcal{S}_\mathrm{hBS}^{\mathrm{imp}}(T=0)=\ln(\frac{1}{\fl{2\gamma}(\fl{2\gamma}-1)}) ,\quad  \gamma>1
\end{equation}
Similarly, for $T\to \infty$
\begin{equation}
    \mathcal{S}_\mathrm{BS}^{\mathrm{imp}}(T\to \infty)=\ln(\frac{1}{\fl{2\gamma}(\fl{2\gamma}+1)}) ,\quad  \gamma>1
\end{equation}

\section*{Numerical Solution of the TBA Equations}

The numerical solution of the TBA equations involves discretizing the rapidity variable $\lambda$ on a finite interval $[-L, L]$ with $N_\lambda$ points, and truncating the infinite hierarchy of auxiliary functions $\{\eta_n(\lambda)\}$ at a maximal string index $n_{\max}$. The discretization is defined by a uniform grid $\lambda_j = -L + j   d\lambda$ for $j = 0, \ldots, N_\lambda - 1$, where the grid spacing is 
\begin{equation}
d\lambda = \frac{2L}{N_\lambda - 1}.
\end{equation}
The chosen cutoff $L = 40$ is motivated by the rapid exponential decay of the convolution kernels $K(\lambda) = \frac{1}{2 \cosh(\pi \lambda)}$ and the driving term $D(\lambda) = -\frac{2 \pi}{T} \frac{1}{\cosh(\pi \lambda)}$, which ensures that contributions for $|\lambda| > L$ are negligible. The discretization size $N_\lambda = 1024$ provides sufficient resolution to accurately evaluate integrals via numerical quadrature and capture the smooth variation of $\eta_n(\lambda)$ and the logarithmic kernels $\ln(1 + \eta_n(\lambda))$.

The infinite set of coupled nonlinear integral equations,
\begin{equation}
\ln \eta_n(\lambda) = \delta_{n, n_{\mathrm{imp}}} D(\lambda) + \int_{-L}^{L} K(\lambda - \lambda') \left[ \ln(1 + \eta_{n-1}(\lambda')) + \ln(1 + \eta_{n+1}(\lambda')) \right] d\lambda',
\quad n = 1, \ldots, n_{\max},
\label{eq:TBA}
\end{equation}
is truncated at $n_{\max} = 30$ and here $D(\lambda) = -\frac{2\pi}{T}\frac{1}{\cosh\left(\pi \lambda\right)}$ is the driving term. The truncation is supplemented by an asymptotic closure condition for $\eta_{n_{\max}+1}(\lambda)$ involving convolution operators defined by
\begin{equation}
[n] f(\lambda) = \int_{-L}^L \frac{n \pi / 2}{\pi \left( (n \pi / 2)^2 + (\lambda - \lambda')^2 \right)} f(\lambda')   d\lambda',
\end{equation}
which ensures physical consistency and proper boundary conditions at large $n$. In the implementation, the closure is imposed pointwise through
\begin{equation}
\log \left(1 + \eta_{n_{\max}+1}(\lambda)\right) = 
\frac{-H/T + [n_{\max}{+}1]\log \left(1 + \eta_{n_{\max}}\right) - [n_{\max}]\log \left(1 + \eta_{n_{\max}-1}\right)}
{[n_{\max}{+}1]1}.
\end{equation}

The numerical solution proceeds by a fixed-point iterative scheme: 
\begin{enumerate}
    \item Initialize \(\eta_0^{(1)}(\lambda) = 0\) and \(\eta_n^{(1)}(\lambda) = (n+1)^{2} - 1\) for \(n \geq 1\) (constant in \(\lambda\)).
    \item For each \(n = 1, \ldots, n_{\max}\), form the right-hand side of Eq.~\eqref{eq:TBA} using \(\eta_{n-1}\) and \(\eta_{n+1}\) from the previous iteration, adding the driving term only when \(n = n_{\mathrm{imp}}\).
    \item Update \(\eta_n(\lambda) \leftarrow \exp \big(\mathrm{rhs}(\lambda)\big)\).
    \item Update \(\eta_{n_{\max}+1}(\lambda)\) from the closure condition.
\end{enumerate}
Convergence is declared once the maximum difference
\begin{equation}
\max_{1 \leq n \leq n_{\max}} \max_{\lambda} \left|\eta_n^{(k+1)}(\lambda) - \eta_n^{(k)}(\lambda)\right| < \mathrm{tol},
\end{equation}
with \(\mathrm{tol} = 10^{-10}\), is achieved.
 This stringent tolerance ensures that numerical errors in the computed $\eta_n(\lambda)$ are minimal, yielding stable and accurate thermodynamic quantities derived from these functions.

\section*{Finite-Temperature MPO Simulation of a Spin-$\frac{1}{2}$ Chain with Impurity}

The density matrix $\rho(\beta)=e^{-\beta H}$ at inverse temperature $\beta$ is represented as a Matrix Product Operator (MPO) and evolved in imaginary time by a Suzuki-Trotter decomposition with two-site gates. We simulate the finite temperature density matrix of a spin-$\frac{1}{2}$ chain of length $N=501$ containing an impurity coupled at site 1 by implementing the purification method~\cite{verstraete2004matrix,zwolak2004mixed} in the ITensor library~\cite{ITensor1,Itensor2}. 

The initial MPO is constructed as the identity operator on the full Hilbert space,
\begin{equation}
\rho(0) = \mathbb{I},
\end{equation}
acting on the Hilbert space $\mathcal{H} = (\mathbb{C}^2)^{\otimes N}$ with dimension
\begin{equation}
\dim \mathcal{H} = 2^{N}.
\end{equation}
This MPO is \emph{not} normalized by $\mathrm{Tr}(\mathbb{I})$. Instead, normalization is managed dynamically by explicitly tracking the trace of $\rho(\beta)$ at each imaginary time step.

Imaginary-time evolution proceeds by applying local two-site gates
\begin{equation}
G_{ij}(\tau) = e^{\tau h_{ij}},
\end{equation}
where $\tau = -\frac{\delta \tau}{2}$, constructed using ITensor’s operator exponentiation. The Hamiltonian $H\sum_{\langle i,j\rangle} h_{ij}, \qquad
h_{ij} = J_{ij} \vec{\sigma}_i \cdot \vec{\sigma}_j,$ is decomposed into nearest-neighbor terms including the impurity bond with coupling $J$. Each Trotter step updates the MPO according to
\begin{equation}
\rho(\beta + \delta \tau) \approx G \rho(\beta) G^\dagger,
\end{equation}
with subsequent compression using a cutoff of $10^{-10}$ to maintain computational efficiency. Here, $
G = \prod_{\langle i,j\rangle} G_{ij}$.

At each inverse temperature $\beta$, the partition function is obtained from the MPO trace,
\begin{equation}
Z(\beta) = \mathrm{Tr}(\rho(\beta)) = \mathrm{Tr}\left(e^{-\beta H}\right),
\end{equation}
which satisfies
\begin{equation}
Z(0) = 2^{N}.
\end{equation}
The total internal energy is computed as
\begin{equation}
E_{\mathrm{tot}}(\beta) = \frac{\mathrm{Tr}(\rho(\beta) H)}{\mathrm{Tr}(\rho(\beta))}.
\end{equation}
The entropy follows from the thermodynamic relation
\begin{equation}
S(\beta) = \beta E_{\mathrm{tot}}(\beta) + \log Z(\beta).
\end{equation}
Dividing by system size yields the energy and entropy per site:
\begin{equation}
e(\beta) = \frac{E_{\mathrm{tot}}(\beta)}{N}, \quad s(\beta) = \frac{S(\beta)}{N} = \beta e(\beta) + \frac{\log Z(\beta)}{N}.
\end{equation}

To isolate the effect of the impurity, a separate MPO imaginary-time evolution is performed for the clean spin-$\frac{1}{2}$ chain of length $N=500$ without the impurity coupling. The clean chain entropy, $S_{\mathrm{bulk}}(\beta)$, is computed analogously. The impurity entropy is then defined as the difference
\begin{equation}
S_{\mathrm{imp}}(\beta) = S(\beta) - S_{\mathrm{bulk}}(\beta),
\end{equation}
quantifying the contribution of the impurity to the thermodynamic entropy.

In the main text, we demonstrate that the results obtained from the numerical solution of the TBA equations and those from the finite-temperature MPS simulations are in excellent agreement.

\section*{Results from TBA and Exact Diagonalization}

\begin{figure}[H]
    \centering
\includegraphics[width=0.6\linewidth]{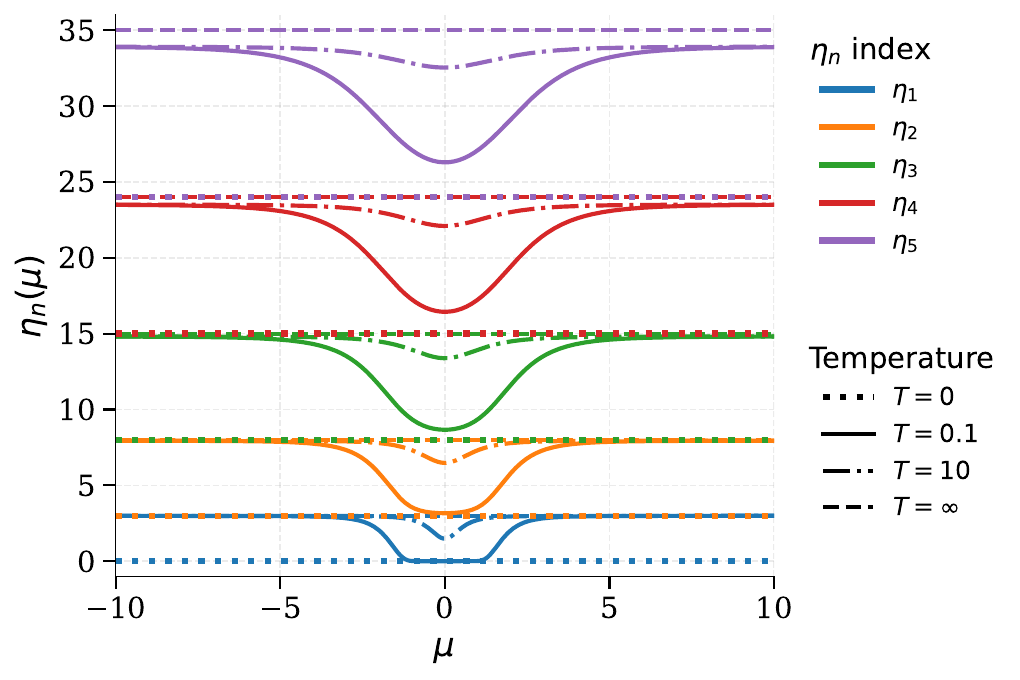}
    \caption{TBA functions $\eta_n(\mu)$ for $n=1,\dots,5$ at 
    temperatures $T=0$, $0.1$, $10$, and $\infty$. Solid and dash–dotted 
    curves correspond to representative finite values of temperatures, while dotted and dashed 
    horizontal lines indicate the zero– and infinite–temperature limits.}
    \label{fig:tba_functions}
\end{figure}

Using the method described above, we compute the string functions 
$\eta_n(T,\mu)$ as functions of the spectral parameter $\mu$ at different 
temperatures. Fig.~\ref{fig:tba_functions} shows representative results 
for $n=1,\dots,5$. For each $n$, the solid and dash–dotted curves correspond 
to representative temperatures $T=0.1$ and $T=10$, while the dotted and dashed 
horizontal lines represent the analytic $T=0$ and $T\to\infty$ limits, 
respectively. As expected, at zero temperature the functions saturate to the 
known constant values $\eta_n(\mu)=n^2-1$, while in the infinite–temperature 
limit they approach $(n+1)^2-1$. At intermediate temperatures, 
$\eta_n(\mu)$ develops a smooth dependence on $\mu$, interpolating between 
these two limits. The variation is most pronounced at small $|\mu|$, while 
for large $|\mu|$ the functions quickly saturate to their asymptotic values.  
This behavior demonstrates that the numerical solution of the thermodynamic Bethe ansatz correctly 
reproduces the expected limiting behavior of the string functions, while 
also capturing their nontrivial dependence on both $\mu$ and $T$. 

After obtaining the functions $\eta_n$, we can proceed to compute the impurity contribution to the 
free energy $F_{\mathrm{imp}}$ by summing over the contributions from the 
different towers of excitations. From this, the corresponding entropy 
contributions of each tower are readily extracted. While the final 
finite-temperature results obtained via tensor-network methods are presented 
in the main text, here we focus on analyzing the separate contributions of the 
individual towers within the antiferromagnetic bound-mode phase. In particular, 
we examine the contributions from two towers in the AFBM phase and compare the outcomes obtained from the thermodynamic Bethe ansatz with those from exact diagonalization.

\begin{figure}[H]
    \centering
\includegraphics[width=0.6\linewidth]{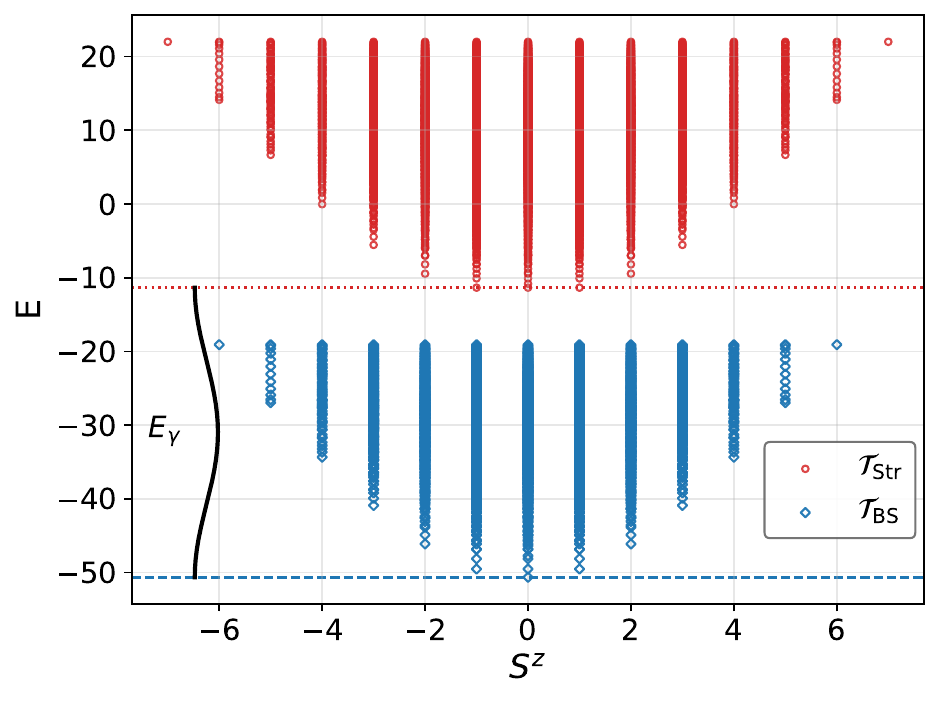}
    \caption{Spectrum of a chain with $N=13$ bulk sites and an impurity 
coupled with $g=1$ and $J=10$. The states are separated into two distinct 
towers: the boundary--string tower $\mathcal{T}_{\mathrm{BS}}$ (blue), 
containing $4096$ states (one quarter of the Hilbert space), and the 
string tower $\mathcal{T}_{\mathrm{str}}$ (red), containing the remaining 
$12288$ states (three-quarters of the Hilbert space), out of a total of 
$16384=2^{14}$ states. The energy scale $E_\gamma$ (black brace) 
separates the two towers. In the infinite--volume limit, 
$E_\gamma = \bigl|2\pi g / \sin(\pi\gamma)\bigr| \approx 39.143$, while 
for the finite chain with $N=13$ bulk sites we obtain 
$E_\gamma^{N=13} \approx 39.290$.}
    \label{fig:N14specJ10}
\end{figure}

In general, for smaller values of the impurity coupling $J$, it is difficult 
to disentangle the two towers within exact diagonalization. To overcome this, 
we employ a strategy in which we first study the problem at large $J$, where 
the two towers are clearly separated. In this regime, we can explicitly verify 
that the boundary–string tower accounts for one quarter of the states, while 
the string tower accounts for the remaining three quarters. As a representative 
example, we show in Fig.~\ref{fig:N14specJ10} the complete spectrum of a 
chain with $N=13$ bulk sites and an impurity coupled to the bulk with $g=1$ 
and $J=10$. Since $E_\gamma$ is large in this limit, the string tower is 
sufficiently lifted, allowing it to be clearly resolved.

\begin{figure}[H]
    \centering   \includegraphics[width=0.65\linewidth]{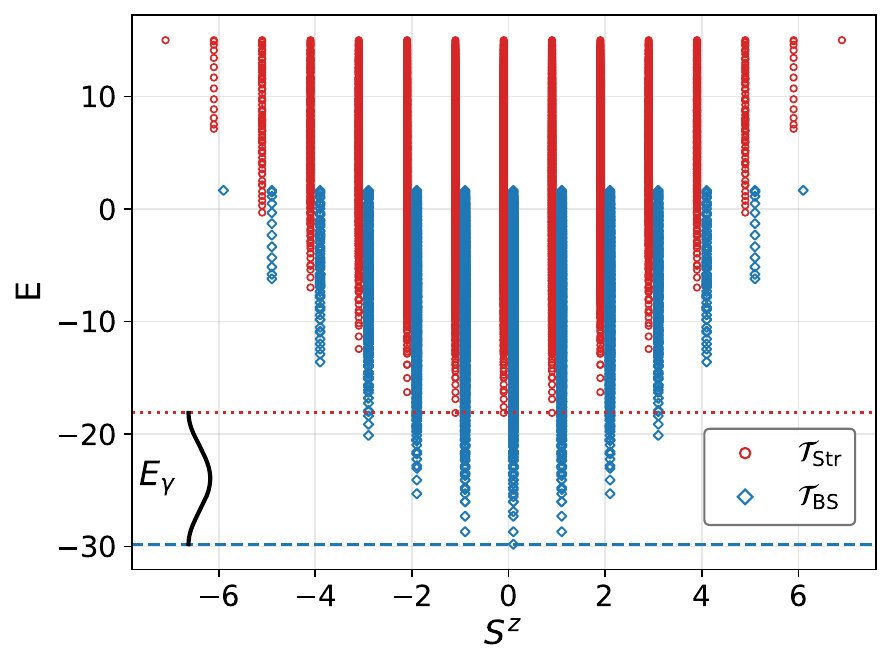}
    \caption{Spectrum of a chain with $N=13$ bulk sites and an impurity 
    coupled with $g=1$ and $J=3$. The two towers, defined by adiabatic 
    continuation from the large--$J_{\rm ref}=10$ reference point, remain well separated 
    by the scale $E_\gamma$ indicated by the brace. In the infinite--volume 
    limit the theoretical prediction is $E_\gamma \approx 11.527$, while for 
    the finite system with $N=13$ bulk sites we obtain 
    $E_\gamma^{N=13} \approx 11.668$.}
    \label{fig:N14specJ3}
\end{figure}

At $J_{\rm ref}=10$ the spectrum exhibits a clean separation into two towers as shown in Fig.\ref{fig:N14specJ10}, and this 
provides a natural reference point for defining the associated subspaces. 
Concretely, we collect the lowest quarter of all eigenstates across the full 
Hilbert space, record their distribution over $S^z$ sectors, and use these to 
construct the rank-$2^{N-2}$ reference projector $P(J_{\mathrm{ref}})$. The 
complementary states define the orthogonal projector $\mathbf{1}-P(J_{\mathrm{ref}})$. 
When the impurity coupling $J$ is reduced, the separation of the two towers 
is no longer visible by eye, but the projector construction can be continued 
adiabatically. For each intermediate value of $J$, the Hamiltonian is 
re–diagonalized in every $S^z$ sector, and the new eigenstates are compared 
to the subspace defined at the previous step. The comparison is made by 
computing the expectation value 
\[
p_n(J)=\langle \psi_n(J) | P(J_{\mathrm{prev}}) | \psi_n(J)\rangle,
\]
which measures the overlap of a given eigenstate with the previously defined 
tower subspace. In each $S^z$ sector, we then retain exactly the same number of 
states as in the reference split, namely the sector counts set at $J=10$, by 
selecting those with the largest overlaps. The updated projector $P(J)$ is the orthogonal projection onto this 
selected set of states. By construction, the procedure maximizes the trace 
$\mathrm{Tr}[P(J_{\mathrm{prev}})P(J)]$, which is precisely the Frobenius–norm 
overlap between successive projectors. In this sense, the protocol provides the 
most adiabatic continuation of the two towers compatible with the $S^z$ 
decomposition, and it allows us to track their evolution reliably down to 
moderate impurity couplings. In what follows, we illustrate the result for the 
representative case $J=3$ in Fig.\ref{fig:N14specJ3}.

\begin{figure}[H]
    \centering    \includegraphics[width=0.65\linewidth]{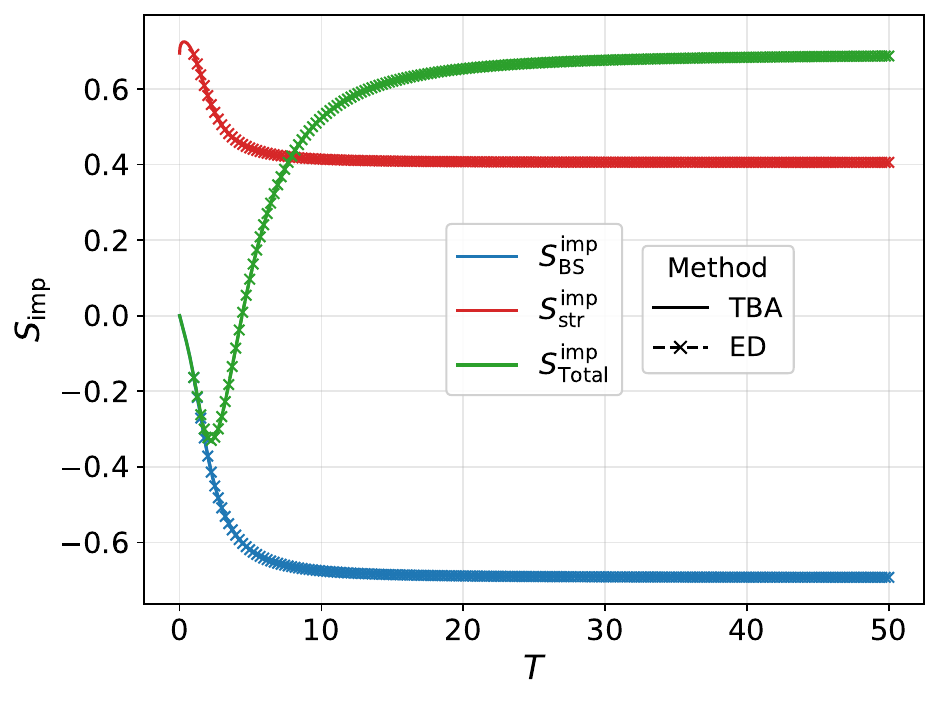}
    \caption{Impurity entropy $S_{\mathrm{imp}}(T)$ at $J=3$, comparing 
    thermodynamic Bethe ansatz (solid lines) with exact diagonalization 
    (crosses). The blue curve shows the contribution from the tower in 
    which the impurity is screened by a bound mode ($\mathcal{T}_{\mathrm{BS}}$), 
    while the red curve corresponds to the tower where the impurity remains 
    unscreened ($\mathcal{T}_{\mathrm{str}}$). The green curve is the total impurity entropy, 
    obtained by combining the two towers with the proper statistical weights, 
    Recalling that the string tower is lifted in energy. The ED results are 
    shown only for $T \gtrsim 1$, where finite--size effects are weak and the agreement with the TBA predictions is excellent.}
    \label{fig:simp3tbaED}
\end{figure}

Having identified the two towers $\mathcal{T}_{\mathrm{BS}}$ and 
$\mathcal{T}_{\mathrm{str}}$, we can compute their thermodynamic 
contributions separately. To do so, we proceed by exact diagonalization of 
the $N=13$ chain with and without the impurity. From the spectrum of the 
pure chain (no impurity) we compute the total entropy $S_{\mathrm{bulk}}(T)$, 
and from the spectrum of the $N=13$ chain with the impurity, we compute the 
total entropy $S_{\mathrm{bulk+imp}}(T)$. The impurity entropy is then 
constructed as 
\[
S_{\mathrm{imp}}(T)  =  S_{\mathrm{bulk+imp}}(T) - S_{\mathrm{bulk}}(T),
\]
and can be decomposed tower by tower according to the adiabatic tracking 
described above. In Fig.~\ref{fig:simp3tbaED} we show the resulting impurity entropy 
from the two towers, comparing the finite--volume results obtained by exact 
diagonalization with the infinite--volume predictions from the TBA. The two 
methods agree very well for $T \gtrsim 1$, while at smaller temperatures the 
ED data exhibit strong finite--size effects

As discussed above, the tower structure is already visible at finite size in the
antiferromagnetic bound–mode (AFBM) regime: the spectrum separates cleanly into two
energetically distinct towers and these can be tracked reliably as $J$ is lowered using the
sector–preserving projector continuation. In contrast, in the ferromagnetic bound–mode
(FBM) and local–moment phases the spectrum organizes into \emph{three} towers. In the
local–moment phase all three towers originate at the same energy, while in the FBM phase
two towers share a common base energy and the third is lifted. Because of these
near– or exact–degeneracies, the overlap–based projector continuation that suffices for two
well–separated subspaces is no longer robust: the per–sector selection becomes ambiguous
and small numerical fluctuations can reshuffle states between the degenerate towers.
Consequently, the exact–diagonalization protocol used in the AFBM regime is not adequate
to disentangle all three towers in FBM or in the local–moment phase.

\begin{figure}[H]
    \centering
    \includegraphics[width=0.65\linewidth]{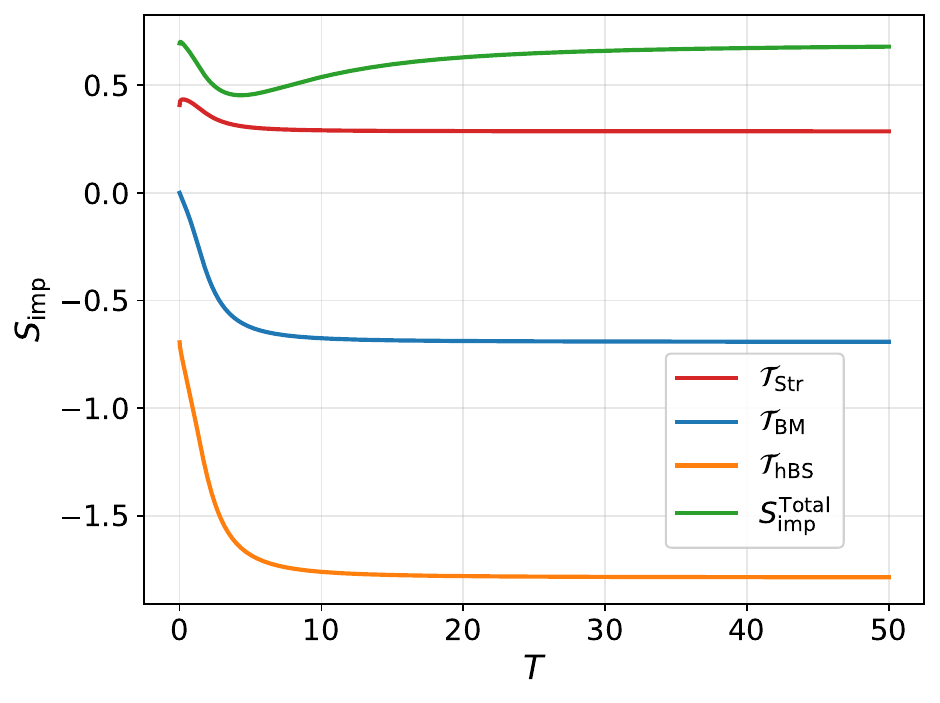}
    \caption{Impurity entropy $S_{\mathrm{imp}}(T)$ at $J=-5$ obtained from the
thermodynamic Bethe ansatz, decomposed into contributions from the three
towers. The red curve corresponds to the string tower $\mathcal{T}_{\mathrm{str}}$, 
the blue curve to the bound--mode tower $\mathcal{T}_{\mathrm{BS}}$, and the 
orange curve to the higher bound--mode tower $\mathcal{T}_{\mathrm{hBS}}$.
Their properly weighted sum gives the total impurity entropy shown in green. 
The low-- and high--temperature limits agree with the analytic predictions: 
$S_{\mathrm{imp}}^{\mathcal{T}_{\mathrm{str}}}(0)=\ln(3/2)$, 
$S_{\mathrm{imp}}^{\mathcal{T}_{\mathrm{BS}}}(0)=0$, 
$S_{\mathrm{imp}}^{\mathcal{T}_{\mathrm{hBS}}}(0)=-\ln 2$, while at $T\to\infty$ one finds
$S_{\mathrm{imp}}^{\mathcal{T}_{\mathrm{str}}}=\ln(4/3)$, 
$S_{\mathrm{imp}}^{\mathcal{T}_{\mathrm{BS}}}=-\ln 2$, and
$S_{\mathrm{imp}}^{\mathcal{T}_{\mathrm{hBS}}}=-\ln 6$.}
    \label{fig:three_towers_Jm5}
\end{figure}

For this reason, we present here a representative infinite–volume calculation based on the thermodynamic Bethe ansatz at $J=-5$, showing separately the contributions to the impurity
entropy from the three towers and their properly weighted total,
$S_{\mathrm{imp}}(T)=\sum_{k=1}^{3} S_{\mathrm{imp}}^{(k)}(T)$
(see Fig.~\ref{fig:three_towers_Jm5}).

\section*{impurity spectral function}
We shall briefly stress that the boundary phase transition is reflected not only in thermodynamic quantities but also in dynamics, transport, and out-of-equilibrium settings. While most of these will be studied in future publications, we shall briefly study the impurity spectral function $A(\omega)$ here, which serves as the definitive fingerprint of how that impurity exchanges spin-flip excitations with its host and whether—and how—it becomes screened.  Because experimental probes such as scanning-tunneling spectroscopy or quantum-dot transport map their differential conductance $dI/dV$ almost one-to-one onto $A(eV)$, a detailed theoretical prediction of $A(\omega)$ allows direct comparison with measurement.

In the present model, four distinct phases arise.  In the Kondo phase, the continuum of bulk spinons with energies ranging from $0$ to $2\pi g$ screens the impurity forming a many-singlet; as we shall see later, in $A(\omega)$ this appears as a broad resonance width is set by the Kondo scale $T_{K}$.  In the antiferromagnetic bound-mode phase, screening is dominated by a single-particle bound mode at energy $E_{\gamma} > 2\pi g$, producing a sharp, delta-like peak at $\omega = E_{\gamma}$ and vanishing weight near zero frequency.  In the ferromagnetic bound-mode phase, the ground state remains four-fold degenerate (the impurity is unscreened at low energy), which shows up as a trivial peak at $\omega=0$, but a high-energy bound-mode resonance again appears at $\omega > E_{\gamma}$.  Finally, in the local-moment phase, the impurity remains unscreened at all energies, and the spectral function exhibits a trivial zero-frequency resonance because of the degeneracy, but there are no bound-state peaks.

To compute $A(\omega)$, we first obtain the ground state $\lvert\Psi_{0}\rangle$
via standard density matrix renormalization group (DMRG) implemented in ITensor library~\cite{ITensor1}.  At $t=0$ we apply the lowering operator $S_{0}^{-}$ to create a local spin-flip,
$\lvert\Phi(0)\rangle = S_{0}^{-}\lvert\Psi_{0}\rangle$,
and then evolve this state under $H$ using the time-dependent variational principle (TDVP) on the manifold of matrix-product states.  This yields the real-time autocorrelation
\begin{equation}
C(t)
= \langle\Psi_{0}\rvert S_{0}^{+} e^{-iHt} S_{0}^{-}\lvert\Psi_{0}\rangle
= \bigl\langle S_{0}^{+}(t) S_{0}^{-}(0)\bigr\rangle,
\end{equation}
sampled at discrete times $0 \le t_{i} \le t_{\max}$.  Choosing a small broadening $\eta \ll J$ to suppress finite-time Gibbs oscillations, we then form
\begin{equation}
G_{\mathrm{imp}}^{R}(\omega)
\approx - i \sum_{i} \Delta t  e^{ i(\omega + i\eta)t_{i}} C(t_{i})
\end{equation}
and extract
\begin{equation}
A(\omega)
= - \frac{1}{\pi} \Im G_{\mathrm{imp}}^{R}(\omega).
\end{equation}

\begin{figure}[H]
  \centering

  \begin{minipage}[b]{0.48\textwidth}
    \centering
    \includegraphics[width=\linewidth]{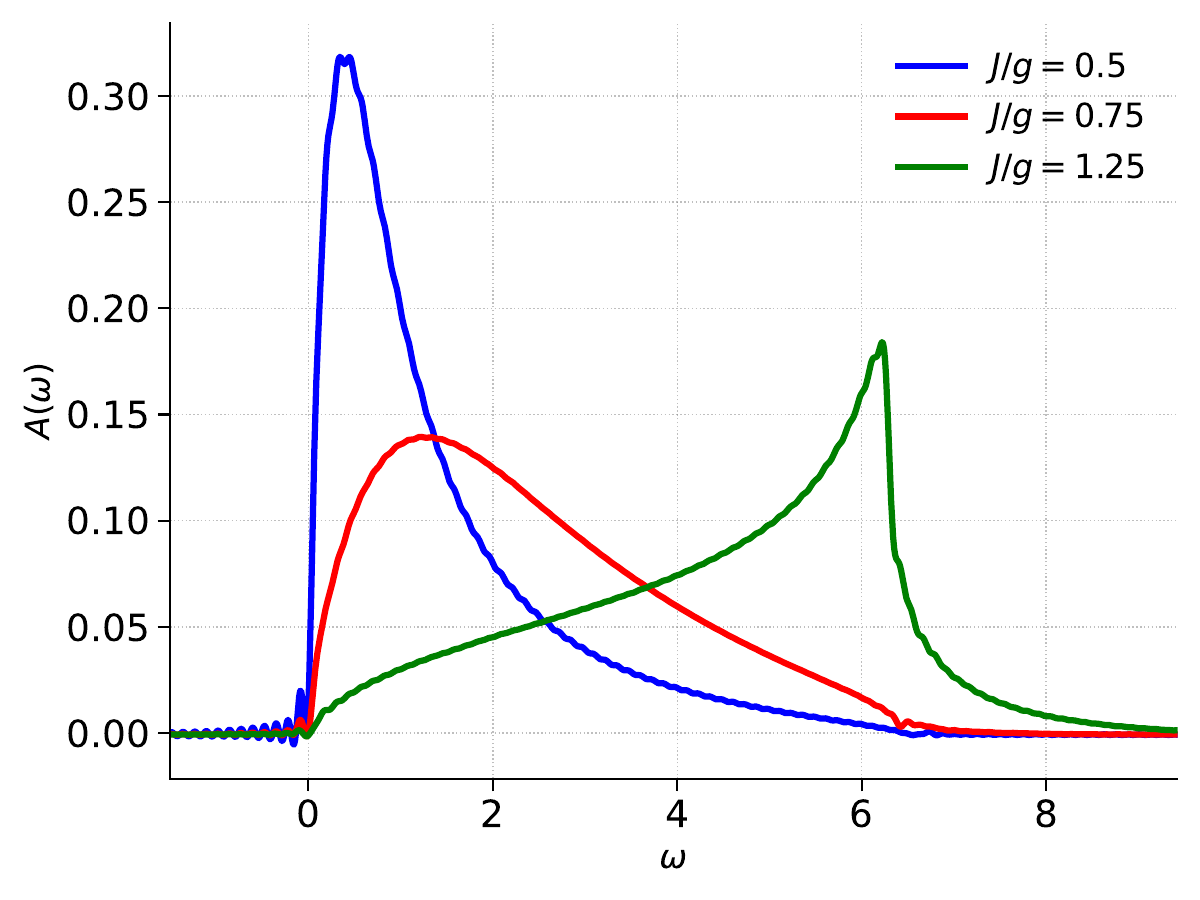}
    \par\vspace{0.4em}
    \textit{(a) Spectral function in the Kondo phase.}
  \end{minipage}\hfill
  \begin{minipage}[b]{0.48\textwidth}
    \centering
    \includegraphics[width=\linewidth]{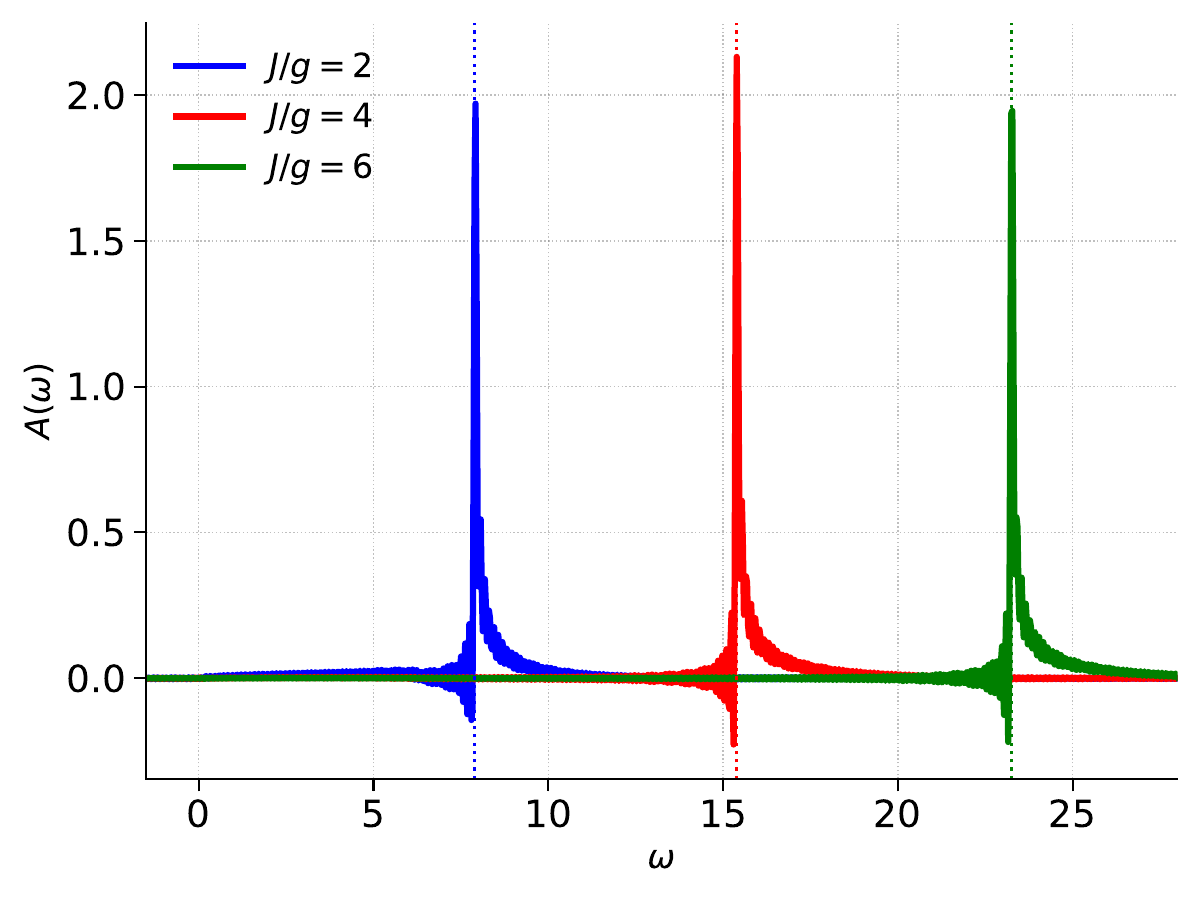}
    \par\vspace{0.4em}
    \textit{(b) Spectral function in the ABM phase.}
  \end{minipage}

  \vspace{1em}

  \begin{minipage}[b]{0.48\textwidth}
    \centering
    \includegraphics[width=\linewidth]{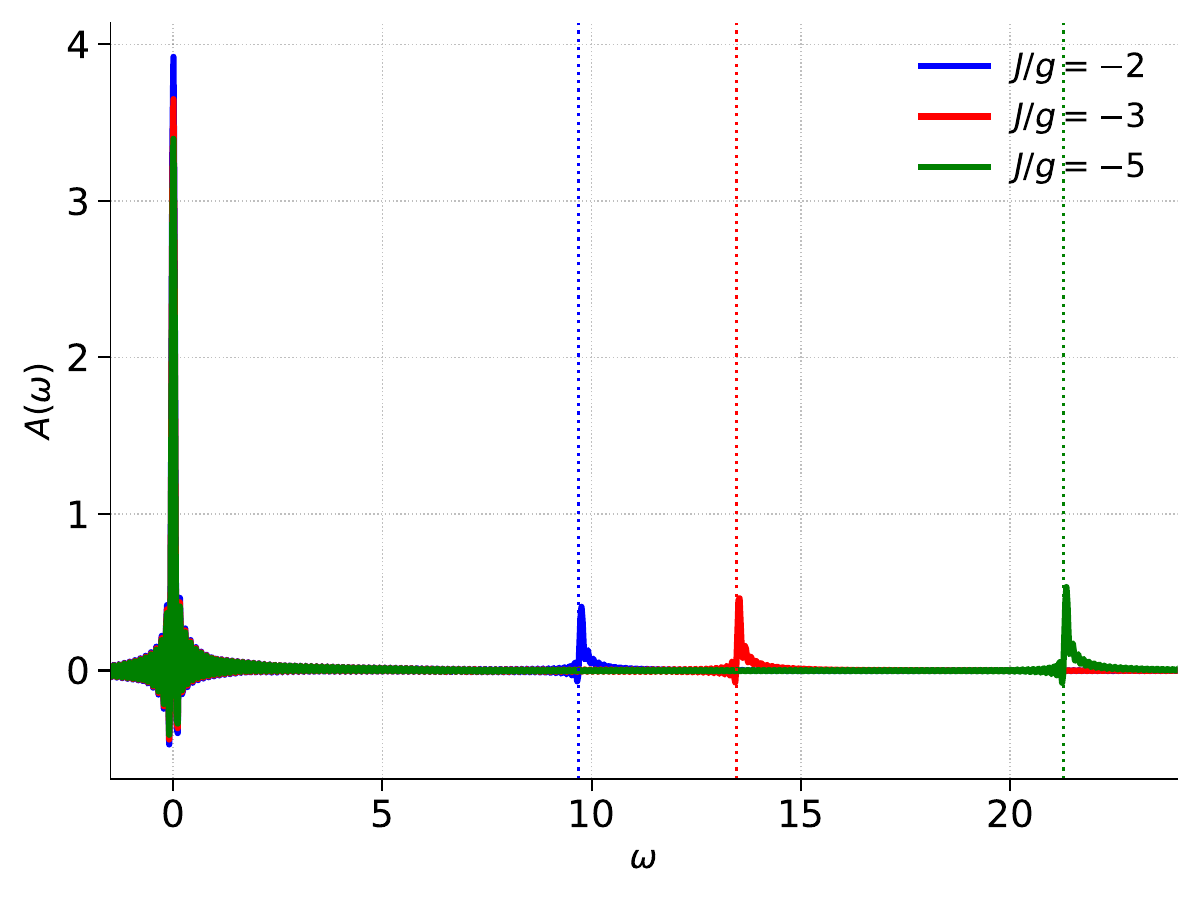}
    \par\vspace{0.4em}
    \textit{(c) Spectral function in the FBM phase.}
  \end{minipage}\hfill
  \begin{minipage}[b]{0.48\textwidth}
    \centering
    \includegraphics[width=\linewidth]{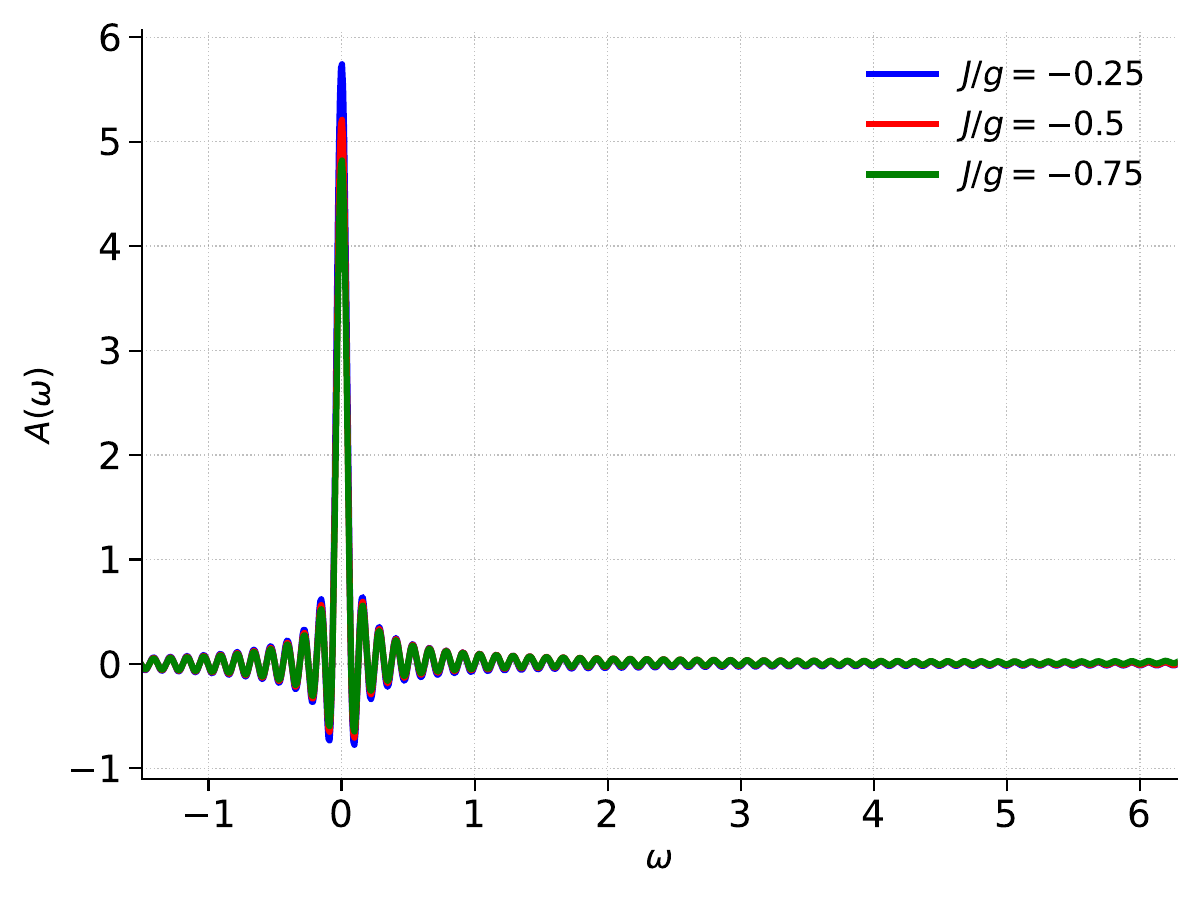}
    \par\vspace{0.4em}
    \textit{(d) Spectral function in the US phase.}
  \end{minipage}

  \caption{Impurity spectral functions $A(\omega)$ for various impurity couplings.}
  \label{fig:impurity-spectra}
\end{figure}

Using a chain of length $N=200$ sites (199 bulk sites and a single impurity at the boundary), we perform TDVP time evolution up to $t=50$ and compute the impurity spectral function via the Fourier transform of the DMRG‐obtained autocorrelation $C(t)=\langle S_0^+(t)S_0^-(0)\rangle$ and extract the spectral function as discussed above.

As shown in Fig.~\ref{fig:impurity-spectra}a), in the Kondo phase, the entire spectral weight lies within the spinon continuum,
\begin{equation}\label{eq:continuum}
0 < \omega = < 2\pi g,
\end{equation}
and for $J\ll g$ one recovers the familiar narrow resonance peaked at $\omega\approx0$.  As $J$ increases, this resonance broadens and steadily shifts toward the upper band edge at $\omega\approx2\pi$, so that just below the phase boundary most of the weight is concentrated at high energy.  This behavior demonstrates that high‐energy spinons, rather than low‐energy ones, dominate the screening cloud as the coupling increases.

In the antiferromagnetic bound‐mode phase Fig.~\ref{fig:impurity-spectra}b), the spectral function collapses onto a single $\delta$‐function peak at the bound‐state energy
\begin{equation}\label{eq:bound}
\omega_{\gamma} =\left|\frac{2\pi g}{\sin(\pi \gamma)}\right|,
\end{equation}
confirming that screening is effectively by a single particle bound mode that is exponentially localized near the edge that hosts the impurity.

Upon entering the ferromagnetic bound‐mode phase under strong ferromagnetic coupling Fig.~\ref{fig:impurity-spectra}c), the ground state remains four‐fold degenerate, and the spectral function exhibits non-vanishing trivial peaks around $\omega=0$
reflecting the degeneracy due to the unscreened impurity, together with a smaller peak at $\omega=\omega_{\gamma}$ indicating partial screening by a high‐energy bound mode.  Finally, in the local‐moment phase Fig.~\ref{fig:impurity-spectra}d), the impurity is unscreened at all energy scales, and $A(\omega)$ shows only the trivial $\omega=0$ degeneracy peak with no additional features.

\end{document}